\documentclass[reprint,twocolumn,amsmath,amssymb,aps,superscriptaddress]{revtex4-2}

\usepackage{graphicx}
\usepackage{dcolumn}
\usepackage{bm}
\usepackage{graphicx}
\usepackage{amssymb}
\usepackage{epstopdf, epsfig}
\usepackage{yfonts}
\usepackage[colorlinks=true, linkcolor=blue, citecolor=blue, urlcolor=blue]{hyperref}
\usepackage{fontenc}
\usepackage{upgreek}
\usepackage{mathrsfs}
\usepackage{booktabs}
\usepackage{pifont}
\usepackage{amsmath}
\usepackage{bm}
\usepackage{wasysym}
\usepackage{caption}
\usepackage{subcaption}
\usepackage{euscript}
\usepackage{natbib}
\usepackage{lipsum}
\usepackage{xcolor}

\usepackage{orcidlink}

\usepackage{subcaption}
\usepackage{ragged2e}
\DeclareCaptionJustification{justified}{\justifying}
\newcommand{\x}[1]{\textcolor{black}{#1}}
\newcommand{\xx}[1]{\textcolor{black}{#1}}

\newcommand{\jkx}[1]{\textcolor{black}{#1}}
\newcommand{\jkxx}[1]{\textcolor{black}{#1}}
\newcommand{\jkv}[1]{\textcolor{black}{#1}}
\newcommand{\jkvv}[1]{\textcolor{black}{#1}}

\newcommand{\vvg}[1]{\textcolor{black}{#1}}
\newcommand{\xxg}[1]{\textcolor{black}{#1}}

\newcommand{\vv}[1]{\textcolor{black}{#1}}
\newcommand{\pp}[1]{\textcolor{black}{#1}}
\newcommand{\px}[1]{\textcolor{black}{#1}}
\newcommand{\pk}[1]{\textcolor{black}{#1}}
\newcommand{\qq}[1]{\textcolor{black}{#1}}
\newcommand{\qs}[1]{\textcolor{black}{#1}}
\newcommand{\qx}[1]{\textcolor{black}{#1}}

\newcommand{\rxx}[1]{\textcolor{black}{#1}}
\newcommand{\ryy}[1]{\textcolor{black}{#1}}
\newcommand{\rr}[1]{\textcolor{black}{#1}}
\newcommand{\rp}[1]{\textcolor{black}{#1}}

\newcommand{\z}[1]{\textcolor{black}{#1}}

\newcommand{\sm}[1]{\textcolor{black}{#1}}
\newcommand{\s}[1]{\textcolor{black}{#1}}

\newcommand{\jfm}[1]{\textcolor{blue}{#1}}

\begin{document}

\title{Non-homogeneous approximation for the kurtosis evolution of shoaling rogue waves}

\author{Saulo Mendes\,\orcidlink{0000-0003-2395-781X}}
\email{saulo.dasilvamendes@unige.ch}
\affiliation{Group of Applied Physics, University of Geneva, 1205 Geneva, Switzerland}
\affiliation{Institute for Environmental Sciences, University of Geneva, 1205 Geneva, Switzerland}

\author{Jérôme Kasparian\,\orcidlink{0000-0003-2398-3882}}
\email{jerome.kasparian@unige.ch}
\affiliation{Group of Applied Physics, University of Geneva, 1205 Geneva, Switzerland}
\affiliation{Institute for Environmental Sciences, University of Geneva, 1205 Geneva, Switzerland}

\begin{abstract}
Bathymetric changes have been experimentally shown to \vv{affect} the occurrence of rogue waves. We recently derived a non-homogeneous correction to the spectral analysis, allowing to describe the evolution of the rogue wave probability over a shoal. Here, we extend this work to the evolution of the excess kurtosis of the surface elevation, that plays a central role in estimating rare event probabilities. Furthermore, we provide an upper bound to the excess kurtosis.
In intermediate and deep water regimes, a shoal does not affect wave steepness nor bandwidth significantly, so that the vertical asymmetry between crests and troughs, the excess kurtosis, and the exceedance probability of wave height stay rather constant. In contrast, in shallower water, a sharp increase in wave steepness increases the vertical asymmetry, resulting in a growth of both the tail of the exceedance probability and the excess kurtosis.
\end{abstract}

\keywords{Non-equilibrium statistics ; Rogue Wave ; Stokes perturbation ; Bathymetry}

\maketitle

\section{INTRODUCTION} 

\jkvv{Ocean wave statistics is \px{at the crossroads of} ocean engineering and physical oceanography. Ocean engineers are commonly concerned with \px{both} short-term and long-term wave statistics \citep{Clauss2002}, \vv{while} the mechanism\vvg{s responsible for the formation} \vv{of} extreme waves is the focus in physical oceanography \citep{Toffoli2015}. The unexpected observation of the so-called rogue waves (also known as freak waves) over the past decades \citep{Haver2004} reignited the cross-disciplinary interest in wave statistics. These waves  seemingly \vv{“}appear from nowhere” \citep{Akhmediev2009x}, and are by statistical definition at least twice \pp{taller than} the significant wave height. From an engineering perspective, the performance of theoretical probability models at the tail of the \ryy{wave height} distribution measure\px{s} the\px{ir} \vv{practical} success \ryy{and applicability to structure dimensioning}.}

\jkvv{Applying the signal processing methods of \citet{Rice1945}, \vv{the bulk of} surface gravity waves were demonstrated to follow a Rayleigh distribution \px{of} heights \citep{Higgins1952}. Nevertheless, the Rayleigh distribution is unsuited to capture the tail of the distribution in real ocean conditions \citep{Forristall1978,Tayfun1980}. On the other hand, nonlinear \pp{theories} and their \vvg{associated} probability distributions are inaccurate in a wide range of \vvg{real} ocean conditions \citep{Ewans2020,Teutsch2020}.} \pp{T}hese difficulties were realized early on, \pp{such that} \rxx{\ryy{an approach based on the} expansion of sums of Gram-Charlier series for a weakly non-Gaussian distribution of the ocean surface} \citep{Higgins1963} has been widely favoured. \vvg{As review\xxg{e}d in \citet{Tayfun2020},} \ryy{the} comput\ryy{ation of} surface elevation, crest \vvg{and wave} height distributions require methodologies that are often computationally burdensome. Naturally, the excess kurtosis became the centre of wave statistics in an attempt to transfer the problem from the probability distribution \vvg{to the cumulant expansion} \citep{Bitner1980,Tayfun1990}. The complexity of water wave solutions led to the \vvg{use} of excess kurtosis as a \ryy{practical} alternative to \pk{the evaluation of statistical} distributions \citep{Marthinsen1992,Janssen2006a}. 

\rxx{Over the past decade, experiments and numerical simulations have been performed to assess the effect of shoaling \ryy{of} irregular waves on the amplification of rogue wave intensity and occurrence \citep{Trulsen2012,Raustol2014,Ma2015,Ducrozet2017,Bolles2019,Chabchoub2019,Adcock2021a}. \citet{Trulsen2020} provid\rr{ed} \ryy{experimental data with} the broadest set of conditions and widest range of \ryy{relative} water depths. As reviewed in \citet{Mendes2022b}, three complementary theoretical models for the wave statistics have emerged, albeit they tend to focus on either surface elevation \citep{Moore2020}, crest height \citep{Adcock2021c} or crest-to-trough height statistics \citep{Mendes2021b}. Although the observed probability of exceedance of rogue waves in the experiments of \citet{Trulsen2020} have been well described by the third model \citep{Mendes2021b}, their observed excess kurtosis \ryy{has not been addressed yet}.}
To \ryy{fill this gap}, \vvg{we} \pk{provide an effective \ryy{extension to the} theory} \pk{\ryy{of} energy density redistribution \citep{Mendes2021b}} \vvg{to} \qs{describe} the \qs{evolution of} kurtosis \pk{of wave trains travelling over a shoal}. \ryy{Because the increase of} the vertical asymmetry between crests and troughs \ryy{is} a key \ryy{ingredient of} the amplification of rogue wave probability \ryy{over} a shoal \ryy{\citep{Tayfun2020,Mendes2021b}}, we \rr{derive} an approximation for \ryy{this asymmetry} as a function of water depth, bandwidth and steepness. \rxx{V}ariations \rxx{in} vertical asymmetry \ryy{in intermediate and deep water regimes (}$k_{p}h >0.5$\ryy{)} \rxx{are too small to affect the} amplifi\rxx{cation of r}ogue waves travelling past a shoal, unless either the \pk{spectrum} \qs{is} significantly broad\px{-banded} ($\nu > 0.5$) \xxg{or the steepness \qs{is} large ($\varepsilon \, \rr{= H_{s}/\lambda} > 1/10$)}. \xxg{Accordingly, \ryy{\rr{the resulting upper bound for} the vertical asymmetry} lead\rr{s} to} an upper bound for \pk{the} excess kurtosis, \ryy{a key information} for \ryy{dimensioning structures as well as} \rxx{\rp{for} wave forecast \ryy{\citep{Janssen2009}}.}

\section{THEORETICAL CONSIDERATIONS}

\jkvv{\pp{W}e \ryy{first} re\rp{call} the main \pk{ideas of the} theory of non-homogeneous analysis of water waves travelling \vv{over} a shoal \citep{Mendes2021b}.}
\jkvv{Given a} velocity potential $\Phi \jkxx{(x,z,t)}$ and surface elevation $\zeta \jkxx{(x,t)}$ \rxx{of waves travelling over a horizontally variable water depth $h(x)$}, the average energy \x{density} \jkvv{evolving over a shoal described by} $h(x) = h_{0} + x \nabla h$ with finite constant slope $\vvg{1/20 \leqslant} | \nabla h | < 1$ \vvg{(see \jfm{figure} \ref{fig:shoal})} \xxg{is expressed as}:
\begin{eqnarray}
\nonumber
\mathscr{E} &=&  \frac{1}{2\lambda} \int_{0}^{\lambda} \Big\{  \Big[  \zeta (x,t) +  h(x) \Big]^{2} - h^{2}(x)  +
\\
&{}& \frac{1}{g} \int_{-h(x)}^{\zeta} \left[ \left( \frac{\partial \Phi}{\partial x} \right)^{2} + \left( \frac{\partial \Phi}{\partial z} \right)^{2} \right] dz \Big\} dx   \quad ,
\label{eq:energy2}
\end{eqnarray}
w\px{ith} \vvg{zero-crossing} \jkvv{wavelength} $\lambda$\rr{,} \z{gravitational acceleration} $g$ \rxx{and we abuse the notation for the projection of the gradient of the depth onto the wave direction $\nabla h \equiv \nabla h \cdot \hat{x} \, \ryy{\equiv \partial h / \partial x}$.} The inhomogeneity of both \px{$\mathscr{E}  (x)$ and $\langle \zeta^{2} \rangle_{t}(x)$} \pk{redistributes energy \rp{among wave heights} and transforms the\rp{ir}} \qs{exceedance} probability\rr{. In the case of an initial} Rayleigh \rr{distribution in region I of \jfm{figure} \ref{fig:shoal}, over and past the shoal \rp{(regions II-V)} the exceedance probability reads:}
\begin{eqnarray}
\mathbb{\qs{P}}_{\alpha,\Gamma}(H>\alpha H_{s}) =  \int_{\alpha}^{+\infty} \frac{4\alpha_{\s{0}}}{\Gamma} \, e^{-2\alpha_{\s{0}}^{2}/\Gamma} \, d\alpha_{\s{0}} = e^{-2\alpha^{2}/\Gamma} \quad ,
\label{eq:Rayexc}
\end{eqnarray}
\ryy{where the correction arises from the evolution of an inhomogenous wave spectrum over the shoal \citep{Mendes2021b}} \px{($\langle \cdot \rangle_{t}$ \px{stands for} temporal averag\px{e)}}:
\begin{figure}
\centering
    \includegraphics[scale=0.42]{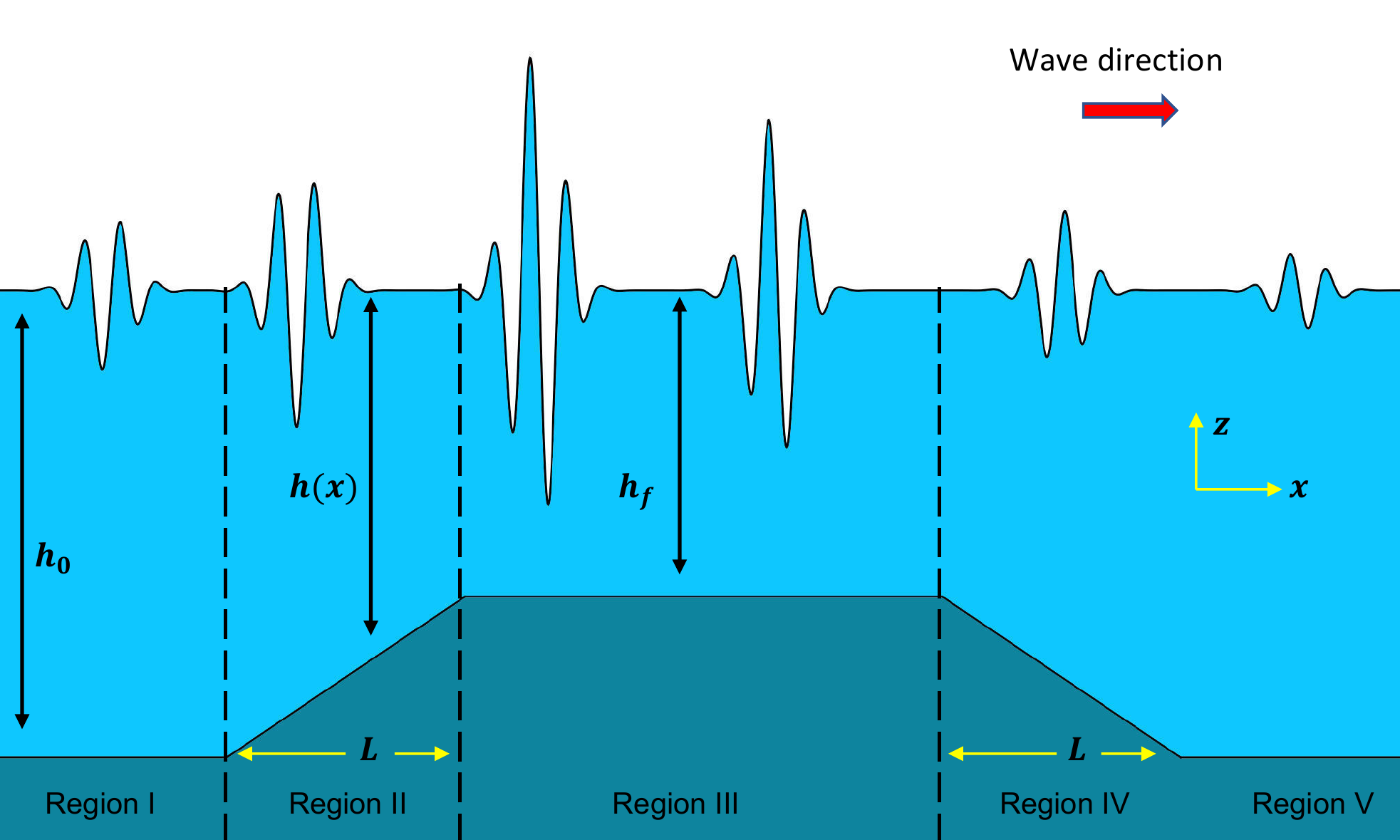}
\caption{Portraying of the extreme wave amplification due to a \vvg{bar} \citep{Mendes2021b}. The water column depth evolves as $h(x) = h_{0} + x \nabla h$ with slope $\nabla h = (h_{f} - h_{0})/L$. Dashed vertical lines delineate shoaling and de-shoaling regions as in \jfm{figure} \ref{fig:modelkurt1}.}
\label{fig:shoal}
\end{figure}
\begin{eqnarray}
\Gamma (x) \approx  \frac{\langle \zeta^{2}(x, t) \rangle_{t} (x)  }{ \mathscr{E}  (x)} \quad \px{.}
\label{eq:GammaEnsemble}
\end{eqnarray}
\ryy{\ryy{The spectral correction} $\Gamma$} depend\ryy{s} on the steepness $\varepsilon = H_{s}/\lambda$ and depth $k_{p}h$, with $H_{s}$ \jkvv{being the} significant wave height\rr{, defined as} \px{the average among the 1/3 largest waves}. \ryy{Note that $H_{s}$ typically differs \rr{by} a few percent from its spectral counterpart $H_{m0}=4\sqrt{m_{0}}$ of Gaussian seas \citep{Holthuijsen2010,Mendes2021a}, where $m_{0}$ is the variance of the surface elevation $\zeta (x,t)$ computed from the wave spectrum. However, this difference can be as large as 10\% in strongly non-Gaussian seas \citep{Goda1983,Mendes2021b}.}
\pk{For linear waves ($\varepsilon \ll 1/100$)\ryy{,} $\Gamma = 1$ and \qs{we} recover the case of a Gaussian sea.} \rxx{When solving $\Gamma$ for second-order irregular waves, we assume \rr{that} th\ryy{e} shoal is linear $(\nabla^{2}h=0)$, the length of the shoal is relatively short $(L/\lambda \lesssim 1)$ and deal with small amplitude waves only ($\zeta/h \ll 1$). These assumptions greatly simplify the problem, but are also representative of real ocean bathymetry \citep{Mendes2022b}. Furthermore, we have recently demonstrated that as the slope magnitude increases the rogue wave occurrence follows suit. \rxx{However, }if we assume small effect of reflection due to a small surf similarity parameter \ryy{among spectral components} \citep{Battjes1974}\ryy{, the increase in rogue wave occurrence} saturates \rr{for slopes larger or equal to} $25^{\circ}$ \citep{Mendes2022b}.} The \px{evolution of the} \pk{exceedance} probability $ \mathbb{P} (H > \alpha  H_{\qs{s}})$ \px{in} eq.~(\ref{eq:Rayexc}) can be \vvg{\px{generalized} to any arbitrary incoming statistics} \citep{Mendes2021b}:
\begin{equation}
\jkx{ \ln  \left( \frac{\mathbb{P}_{\alpha  \, ,\,  \Gamma_{\mathfrak{S}}}}{\mathbb{P}_{\alpha}} \right)  \approx 2\alpha^{2} \left( 1 - \frac{1}{\mathfrak{S}^{2}\qq{(\alpha)}\Gamma_{\mathfrak{S}}}   \right)     \quad , }
\label{eq:bath}
\end{equation}
\vvg{with} the vertical asymmetry between crests and troughs \ryy{being \rr{defined} as twice the \rr{ratio} between crest and crest-to-trough heights}
\citep{Mendes2021a},
\begin{equation}
\ryy{ \mathfrak{S} = \frac{2\mathcal{Z}_{c}}{H} \quad  \therefore \quad    1 \leqslant \mathfrak{S} \leqslant 2 \quad ,}
\label{eq:betalpha20}
\end{equation}
\vvg{\ryy{which for} rogue waves \ryy{features the mean \rr{empirical} value:}}
\begin{eqnarray}
 \mathfrak{S} \xx{(\alpha = 2)}   \approx \frac{\sm{2}\eta_{\vvg{s}}}{1+\eta_{\vvg{s}}} \Bigg( 1 + \frac{ \eta_{\vvg{s}}  }{6}  \Bigg)  \,\, ,
\label{eq:betalpha2}
\end{eqnarray}
\vv{where $\eta_{s}$ measures the ratio between mean crests and mean troughs \ryy{and} has empirically \ryy{been} found \rr{in a wide range of sea conditions} to depend on the skewness \ryy{of the surface elevation} \ryy{$\mu_{3}$} \citep{Mendes2021a}:
\begin{equation}
\eta_{s} \approx 1 + \mu_{3} \quad .
\label{eq:betalpha22}
\end{equation}
\rxx{\ryy{The empirical \rp{relations} of eqs.~(\ref{eq:betalpha2},\ref{eq:betalpha22}) stem from} field observations \ryy{during North Sea storms} d\ryy{etailed} in \jfm{section} \ref{sec:VAfinite}}.} When the water depth decrease\pk{s} waves \pk{become} steeper while the super-harmonic contribution has an increasing share of the wave envelope. The combination of these two effects redistribut\qq{es} the exceedance probability by \qx{causing the rise in} $\langle \zeta^{2} \rangle$ \qx{to exceed the growth of} $\mathscr{E}$. \qx{Such uneven growth} \qq{explains} why a shoal in intermediate water amplifies rogue wave occurrence as compared to deep water \citep{Trulsen2020,Chabchoub2021} while it \pk{reduces this} occurrence in shallow water \citep{Glukhovskii1966,Karmpadakis2022}. 
\begin{figure*}[t]
\minipage{0.48\textwidth}
    \includegraphics[scale=0.47]{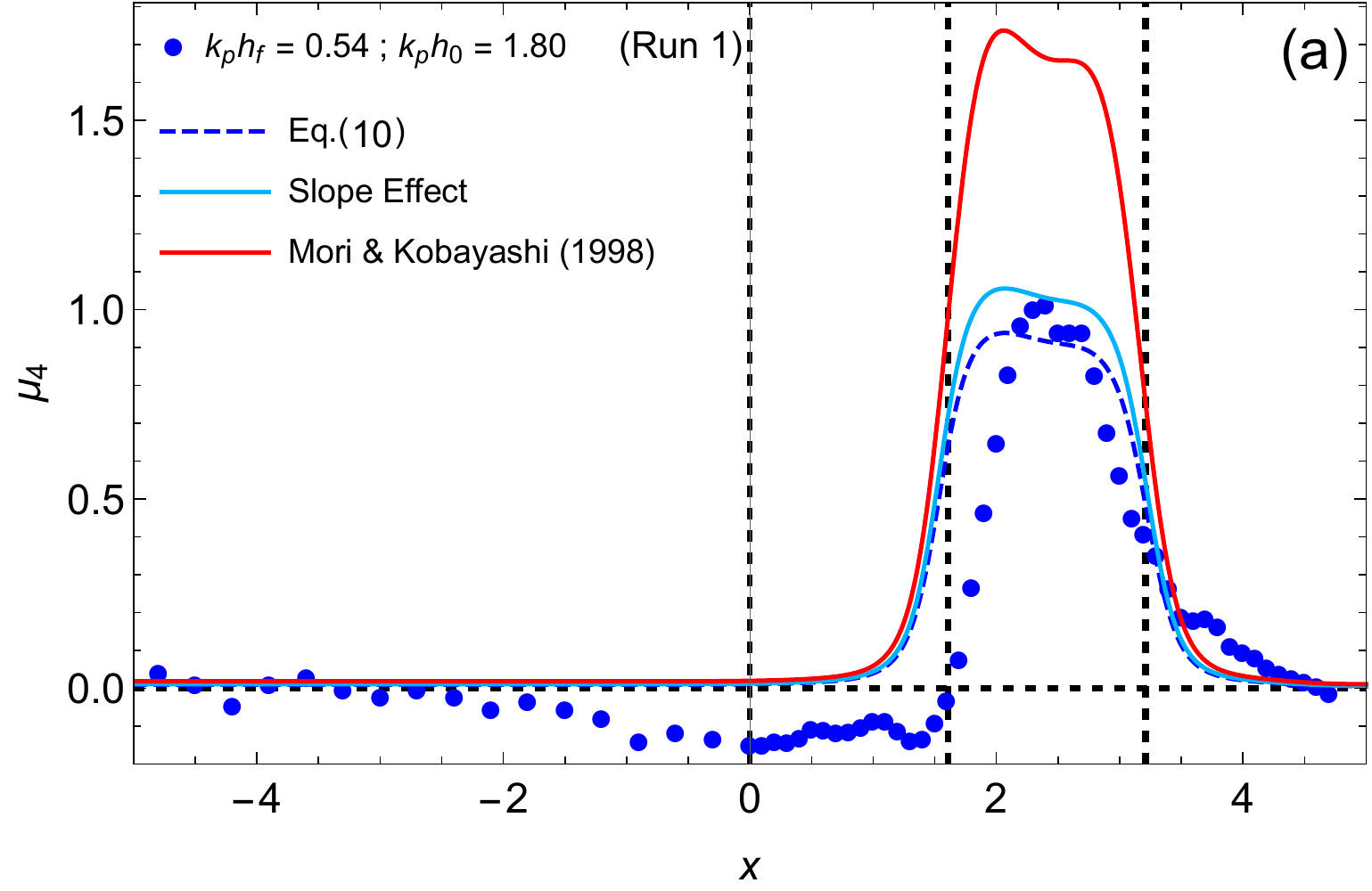}
\endminipage
\hfill
\minipage{0.52\textwidth}
    \includegraphics[scale=0.47]{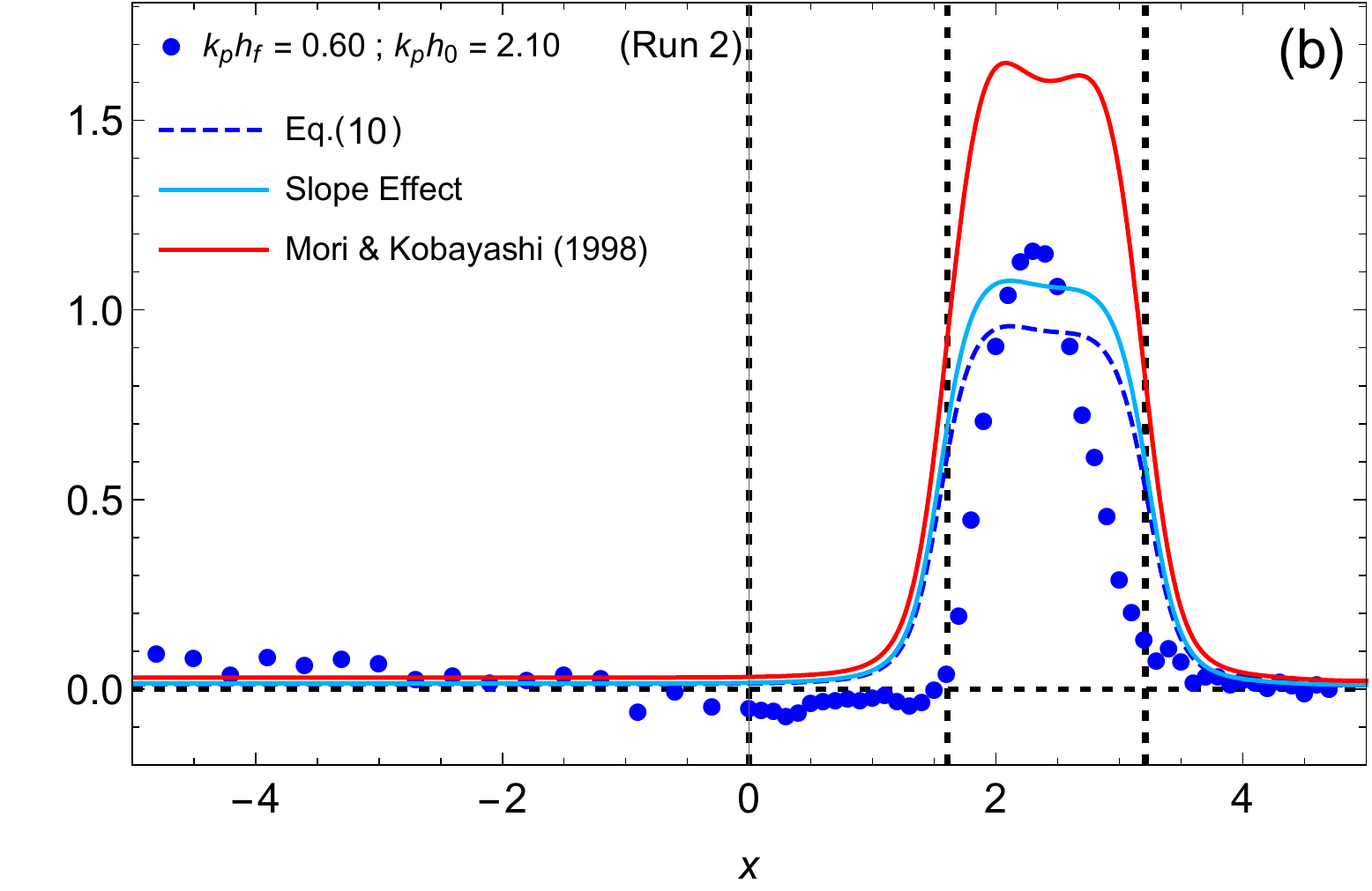}
\endminipage

\minipage{0.48\textwidth}
    \includegraphics[scale=0.46]{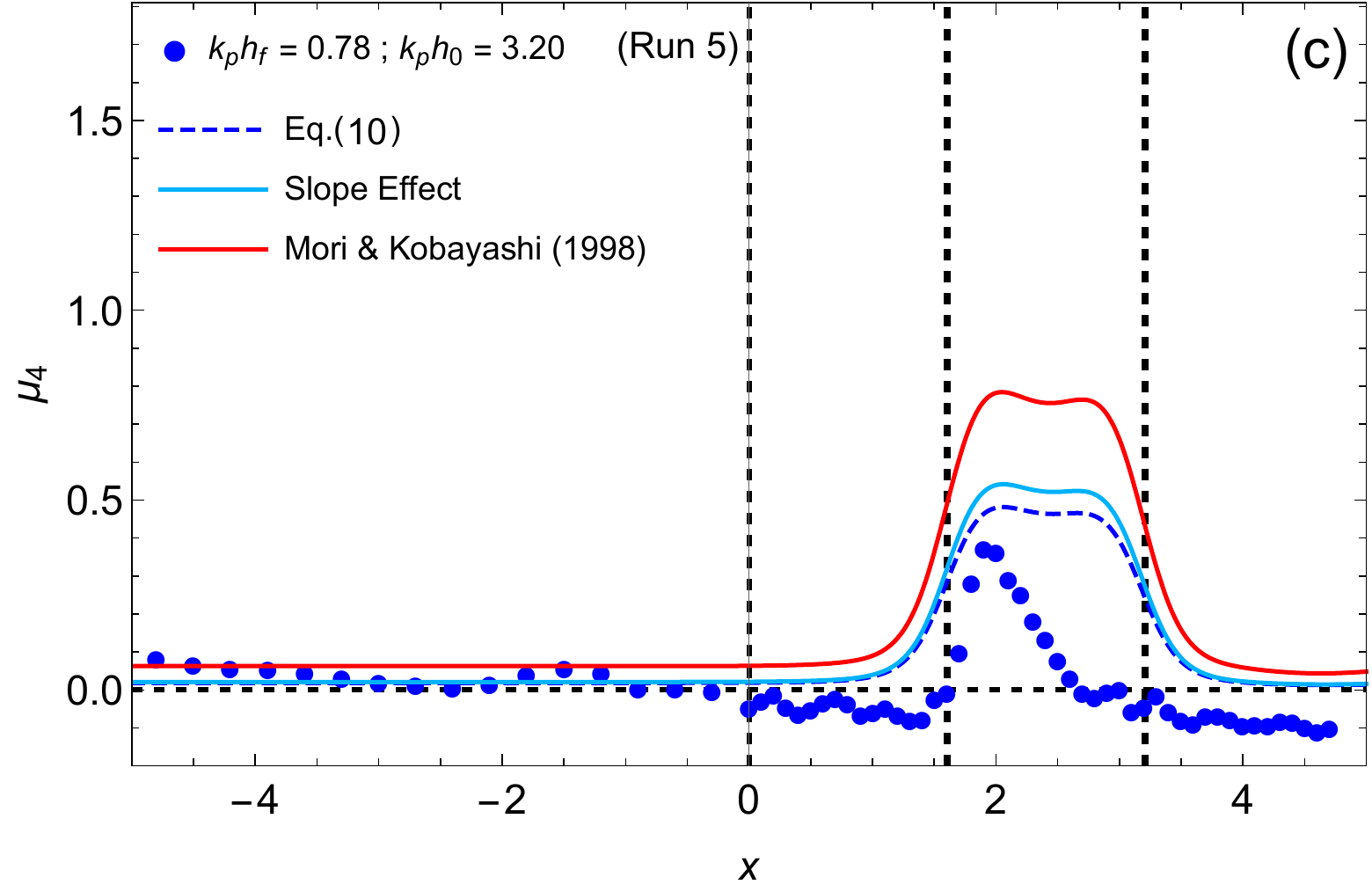}
\endminipage
\hfill
\minipage{0.52\textwidth}
    \includegraphics[scale=0.46]{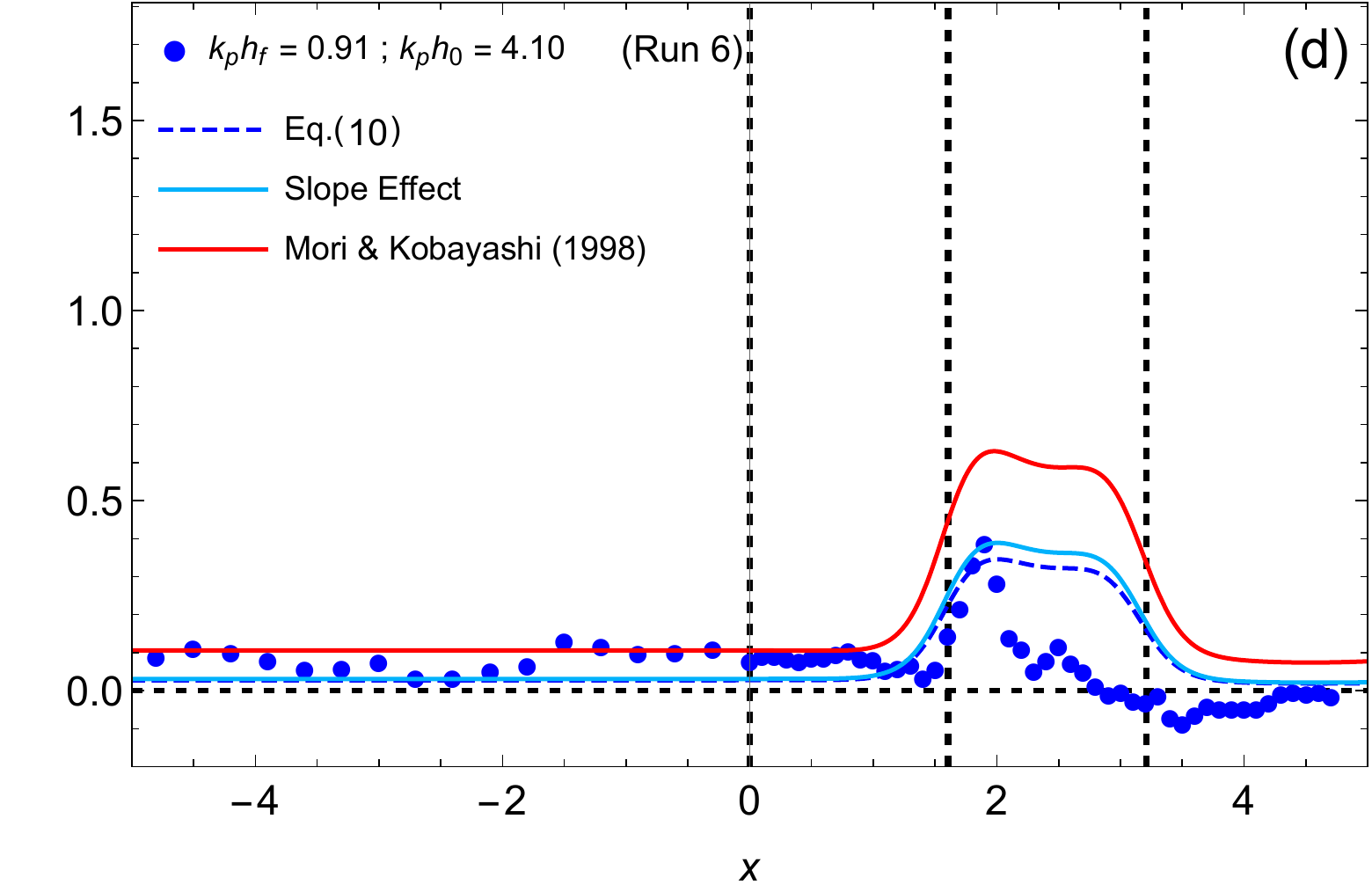}
\endminipage
\caption{Observed kurtosis $\mu_{4}$ (dots) versus the model of eq.~(\ref{eq:Mori3X}) (dashed) for Runs \vv{1, 2, 5, and 6} in \citet{Trulsen2020}. Dashed vertical lines mark the shoaling and de-shoaling zones (see \jfm{figure} \ref{fig:shoal}). \vvg{\ryy{The cyan s}olid curve \qq{include\ryy{s}} the slope effect \citep{Mendes2022b} \ryy{while the red solid curve shows the bound wave prediction according to \citet{Mori1998}.}}}
\label{fig:modelkurt1}
\end{figure*}
The linear term in $\zeta (x,t)$ has the leading order in deep water and $\Gamma - 1 \qs{\lesssim 10^{-2}}$ is \pk{small}\qs{. Conversely,} in intermediate water the super-harmonic creates significant disturbances in the energy density \qq{increasing $\Gamma - 1$ \qx{up to $10^{-1}$}}, whereas in shallow water the super-harmonic diverges and $\Gamma - 1 \qs{\lesssim 10^{-3}}$ becomes \pk{small} again, \qq{reading even} smaller \qq{values} than in deep water.

\section{KURTOSIS EVOLUTION OVER A SHOAL}\label{sec:kurtosis}

The probability evolution of \vvg{eq.~(\ref{eq:Rayexc})} depend\px{s solely} on $\Gamma$. \pk{\ryy{A}ny deviation from a Gaussian distribution may} be described by a \vv{cumulant expansion} \citep{Higgins1963} which \qq{at leading order is expressed as} a function of the excess kurtosis $\vv{\mu_{4}}$. \pk{For the case of an inhomogeneous wave field due to a shoal, there is an excess in kurtosis due to} \px{the} energy partition. \ryy{To avoid the tedious algebra of} eqs. (C1,C7b\ryy{,C12}) of \citet{Mendes2021b} for the case of a non-Gaussian sea prior to the shoal\ryy{, we consider t}he probability \xxg{ratio} relative to the Rayleigh distribution \pk{(implying a pre-shoal $\mu_{4} = 0$) \ryy{to obtain the excess kurtosis}.} \ryy{The ratio measures the amplification of the exceedance probability of waves with height $H = \alpha H_{s}$ due to a shoal and} \px{is computed through the} transformation of variables \px{from the wave envelope in} \citet{Mori2002b} into normalized heights \px{to leading order in $\mu_{4}$\ryy{, as} computed in} section 6.2.3 of \citet{BMendes2020}:
\begin{eqnarray}
\qq{  \frac{\mathbb{P}_{\alpha , \mu_{4}} }{ \mathbb{P}_{\alpha}  } \, \approx \, 1 + \mu_{4}  \frac{\alpha^{2}}{2}  \left( \alpha^{2} - 1  \right)  + \mu_{3}^{2}  \frac{5\alpha^{2}}{18}  \left( 2\alpha^{4} - 6\alpha^{2} - 3  \right)  ,}
\label{eq:Mori1X}
\end{eqnarray}
Taking into account the \qq{theoretical} relation $\mu_{4} \approx 16\mu_{3}^{2}/9$ \px{between kurtosis and skewness} \ryy{for waves of second-order in steepnes} \qq{confirmed by wave shoaling experiments} \citep{Mori1998}, \qq{we rewrite eq.~(\ref{eq:Mori1X}):}
\begin{eqnarray}
\frac{\mathbb{P}_{\alpha , \mu_{4}} }{ \mathbb{P}_{\alpha}  } \, \pp{\approx} \, \pp{1 + \mu_{4} \cdot    \frac{\alpha^{2}}{32} \left( 10\alpha^{4} - 14 \alpha^{2} - 31  \right) } \,\, \qq{, \,\, \forall \, \alpha \gtrsim 2}  \quad .
\label{eq:Mori2X}
\end{eqnarray}
The kurtosis measure\px{s} taildness and \px{it affects the \qq{exceedance} probability \qs{for} $\alpha \gtrsim 1.5$}. \qx{E}q\qs{s}.~(\ref{eq:bath}) \pk{and} (\ref{eq:Mori2X}) \ryy{both} describe the \rr{same consequence} of energy redistribution and the associated deviation from a Gaussian sea\rr{, but the former embodies the physics of shoaling while the latter delineates the perturbation on the statistics regardless of the physical mechanism}. \rr{Therefore, t}hey can be matched, \qx{yielding} a kurtosis $\mu_{4} (\Gamma , \alpha)$. \ryy{This matching could be performed at any value $\alpha \geqslant 1.5$, however, higher accuracy is obtained in the region of stability of the approximation ($2 \lesssim \alpha  \lesssim 3$)\rr{\rp{. Over this range, the resulting value of $\mu_{4}$ deviates by less than 20\%.}}} The\qs{refore,} we \ryy{match} both equations at $\alpha =2$ \ryy{without substantial loss in precision}:
\begin{eqnarray}
\vspace{+0.3cm}
 \mu_{4} (\Gamma) \approx \frac{1}{9} \left[ e^{ 8 \left( 1 - \frac{ 1 }{ \mathfrak{S}^{2}\Gamma } \right) }- 1 \right]     \quad .
\label{eq:Mori3X}
\end{eqnarray}
\qs{This expression generalizes the result obtained by} eqs. 46-47 \qs{of} \citet{Janssen2006a} \qs{in the case of} a narrow-banded wave train, \qs{with less than} 5\% \qs{deviation as compared to their model with a} $(2/3)\alpha^{2}(\alpha^{2} - 1)$ \qs{polynomial in \qx{the counterpart of} eq.~(\ref{eq:Mori2X}) \ryy{for small values of the skewness ($\mu_{3} \ll 1$). However, if the surface elevation is significantly skewed ($\mu_{3} \gtrsim 1$) the contribution of the skewness is severely underpredicted \rr{by eqs. 46-47 of \citet{Janssen2006a}} and therefore the excess kurtosis will be overpredicted while describing the ratio $\mathbb{P}_{\alpha , \mu}/ \mathbb{P}_{\alpha}$}.} \rxx{In order to validate our effective theory for steep slopes of eq.~(\ref{eq:Mori3X}), \ryy{\jfm{figure} \ref{fig:modelkurt1}} compare\ryy{s} \ryy{\rr{its prediction} with} the observed excess kurtosis in \citet{Trulsen2020}.} \qs{In the comparison, w}e employed the empirical \citep{Mendes2021a} asymmetry $\mathfrak{S} (\alpha = 2) = \rxx{1.2}$. \rr{W}e \rr{shall} validate this approximation \rr{i}n the next section \rr{in relative water depth} $k_{p}h \gtrsim \pi/10$, bandwidth $\nu \lesssim 1/2$ \rr{as defined in \citet{Higgins1975}} and \rr{steepness} $\varepsilon \qs{\ll 1/10}$ representative of \citeauthor{Trulsen2020}'s experiments. \rxx{In these experiments, irregular waves with broad-banded JONSWAP spectrum of $\gamma = 3.3$ peak enhancement factor, significant wave height $1.4 \, \textrm{cm} < H_{s} < 3.4 \, \textrm{cm}  $ and peak period $0.7 \, \textrm{s} < T_{p} < 1.1 \, \textrm{s}$ were generated in a 24.6 m long and 0.5 m wide unidirectional wave tank. These irregular waves travelled over a flat bottom that had initial \ryy{relative} water depth ranging from $k_{p}h=4.9$ \ryy{(}deep water\ryy{)} to $k_{p}h=1.8$ \ryy{(}intermediate water\ryy{)}. Furthermore, the irregular waves \ryy{propagated over} a symmetrical breakwater as \rr{sketched} in \jfm{figure} \ref{fig:shoal} with slope $|\nabla h| \approx 1/3.8$ \ryy{on each side and} located 10.8 m after the wavemaker, or equivalently half a dozen peak wavelengths. The \ryy{relative} water depths atop the shoal are in the range $0.54 \leqslant k_{p}h \leqslant 1.60$. \ryy{In addition, the absolute water depths ranged from 0.5-0.6 m prior to the shoal and from 0.08-0.18 m atop the shoal.}} \ryy{E}q.~(\ref{eq:Mori3X}) \ryy{reproduces} well \ryy{the magnitude and} the trend of the \px{peak in} excess kurtosis to decrease towards deep\px{er} waters \xxg{of the experiments in \citet{Trulsen2020}} (see \jfm{figure} \ref{fig:modelkurt1}). \ryy{Remaining differences such as the slightly earlier} rise \ryy{of kurtosis in the shoaling zone and the later fall in the de-shoaling zone are likely due to the assumption of \rp{negligible} reflection.}
\rxx{\ryy{W}e also computed the kurtosis contribution due to the bound wave \ryy{following} \citet{Mori1998} \rr{to evaluate its performance over abrupt changes in relative water depth, see} \ryy{\jfm{appendix} \ref{sec:AppB}}. \rr{Th\rp{is} bound kurtosis model (red curve in \jfm{figure} \ref{fig:modelkurt1}) captures the qualitative trend for the observed kurtosis evolution. However, since the latter was developed for a flat bottom and has no explicit slope dependence \rp{it overestimates the magnitude of the effect. Furthermore}}, our model \rr{in eq.~(\ref{eq:Mori3X})} \ryy{has the advantage of being extendable} to any arbitrary slope \citep{Mendes2022b}.}

\section{VERTICAL ASYMMETRY IN FINITE DEPTH}\label{sec:VAfinite}

\begin{figure}
\begin{subfigure}[t]{0.4\textwidth}
    \includegraphics[scale=0.44]{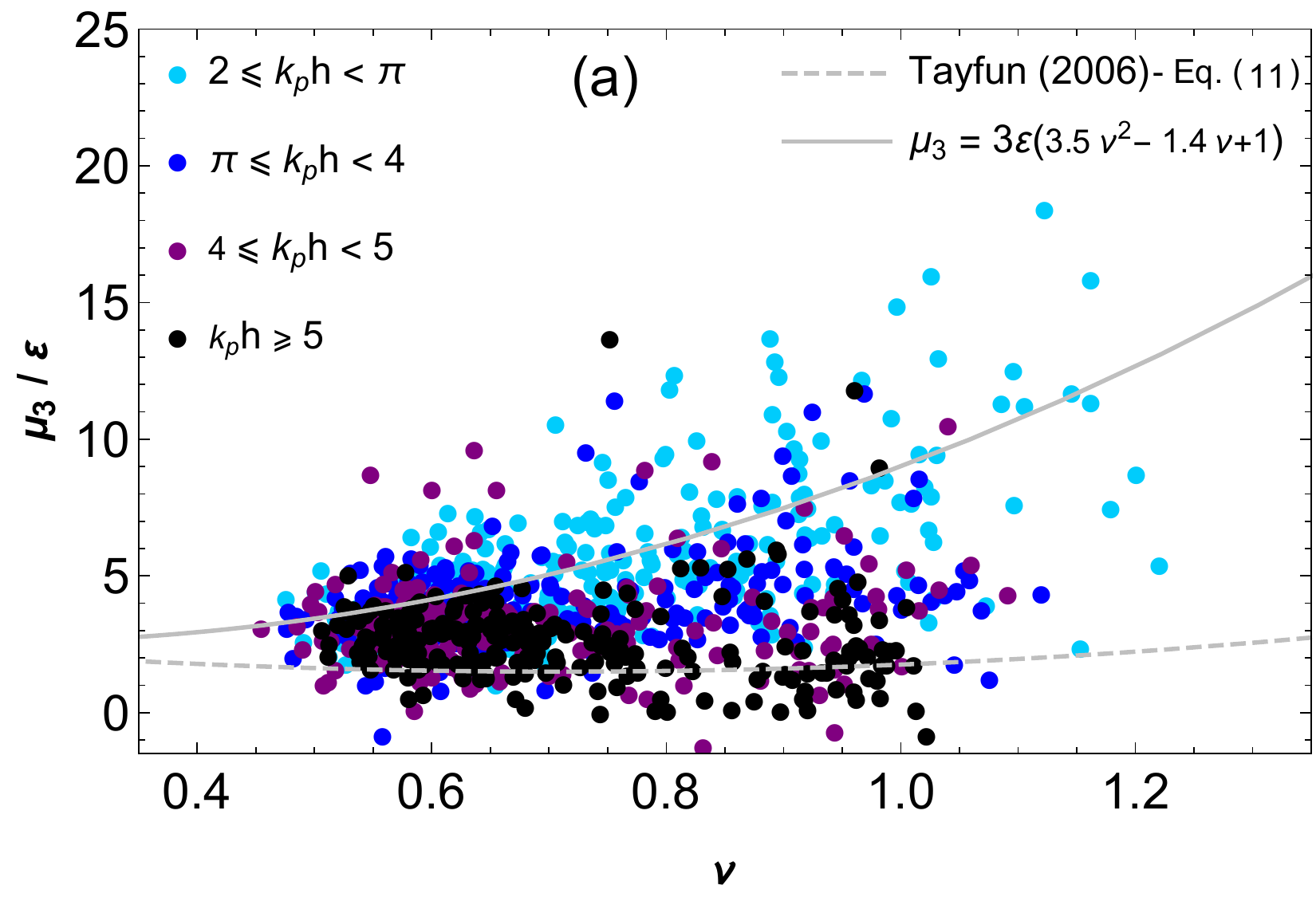}
\end{subfigure}
\hfill
\begin{subfigure}[t]{0.45\textwidth}
    \includegraphics[scale=0.5]{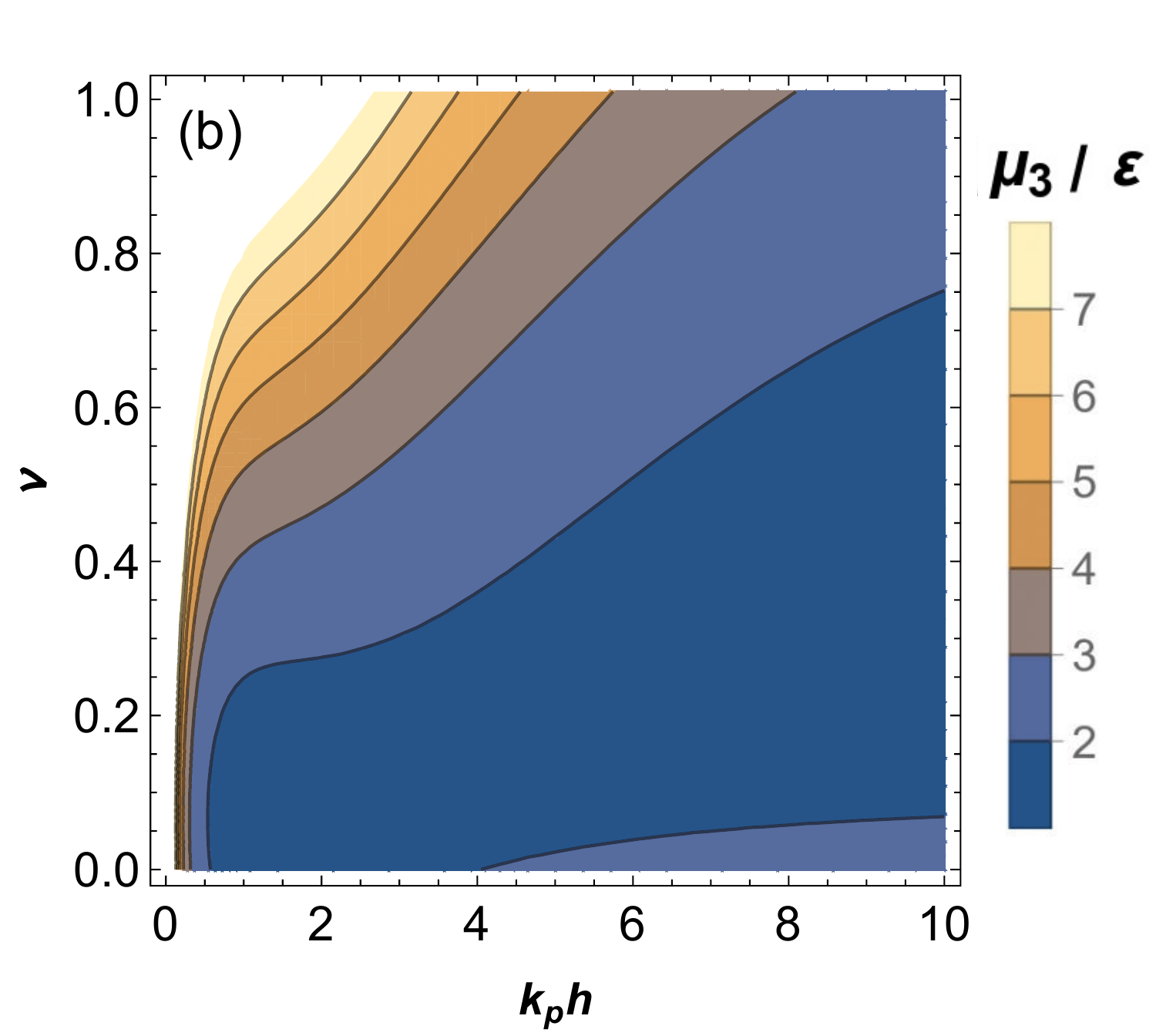}
\end{subfigure}
\caption{\vvg{\pp{(a)} Ratio of skewness and steepness \pk{varying with bandwidth in strongly non-Gaussian ($ \mu_{4} \approx 0.4$) North Sea data \ryy{\citep{Stansell2004}}}, \pk{with polynomial fit $\mathfrak{B}(\nu) \approx 1- \nu \sqrt{2}  + \ryy{3.5} \nu^{2}$ at $2 \leqslant k_{p}h \leqslant \pi$}. (b) \pk{Contour plot of the same ratio as computed from eq.~(\ref{eq:Gnew0}) for \qx{the} fitted \qs{function} $\mathfrak{B}(\nu , k_{p}h)$ \qs{in (a)}.}}}
\label{fig:mu3steep}
\end{figure}

\qs{E}qs.~(\ref{eq:bath}) and (\ref{eq:Mori3X}) \qs{highlight the \qx{influence}} of the vertical asymmetry \qs{on} \qx{the evolution of} rogue wave \qx{occurrence} and excess kurtosis \ryy{of the surface elevation} over a shoal in intermediate depths. However, the evolution of this asymmetry due to finite depth effects \qs{is} not well-known, except that \rp{it} is a slowly varying function \qq{of} the steepness \ryy{\citep{Tayfun2006,Tayfun2020}}. \rxx{To \ryy{\rr{describe} the change in vertical asymmetry due to bandwidth and relative water depth}, we \rr{assess data from} North Sea observations. Data was collected \ryy{on} Total Oil Marine's oil platform \ryy{North Alwyn NAA} located at $60^{\circ} 48.5'$ N and $1^{\circ} 44.2'$ E, approximately 135 km east of the Shetland Islands (Scotland) and 156 km west of the Norwegian coast \citep{Stansell2004,Stansell2005}. The platform sits on a depth of 129 m and \ryy{on} a mild slope of $\nabla h \sim \rp{-} 1/300$ in the SE-NW direction (\ryy{according to} bathymetry charts from \href{https://portal.emodnet-bathymetry.eu/?menu=19}{EMODnet} - European Marine Observation and Data Network, see \jfm{figure} \ref{fig:sketch})\rp{. While} the mean wave direction during the winter storms observed between 1995-1999 \citep{Stansell2000} \rp{is in the SE-NW direction, w}\rr{e focused on the shoaling case, i.e. waves coming from the southeast towards the northwest.} The mild slope is almost linear ($\nabla^{2}h \approx 0$) within a distance of 250 m Northwest and Southeast of the platform, \ryy{corresponding to} three mean wavelengths (see table 3 of \citet{Mendes2021a} \ryy{for the measurements}). The raw data were stored as 2381 20-min records of surface elevation measurements recorded with a sampling rate of 5 Hz.} \ryy{To perform the comparison with ocean data, we f}ollow \citet{Marthinsen1992} \ryy{and} \qs{consider the skewness \ryy{of the surface elevation} to depend solely on \ryy{relative water} depth and \ryy{wave} steepness} $\mu_{3} = \mu_{3} (\varepsilon , k_{p}h )$, and consequently \pk{identify} $ \mathfrak{S}(\mu_{3}) \xxg{= \mathfrak{S}(\varepsilon , k_{p}h)}$ \px{for any $\alpha$} \ryy{due to eq.~(\ref{eq:betalpha2})}. \ryy{We} approximate the skewness as (see eq. 19 of \citet{Tayfun2006}, \ryy{where} $\mu$ denot\ryy{es} steepness and $\lambda_{3}$ the \qs{skewness}):
\begin{figure}
    \includegraphics[scale=0.46]{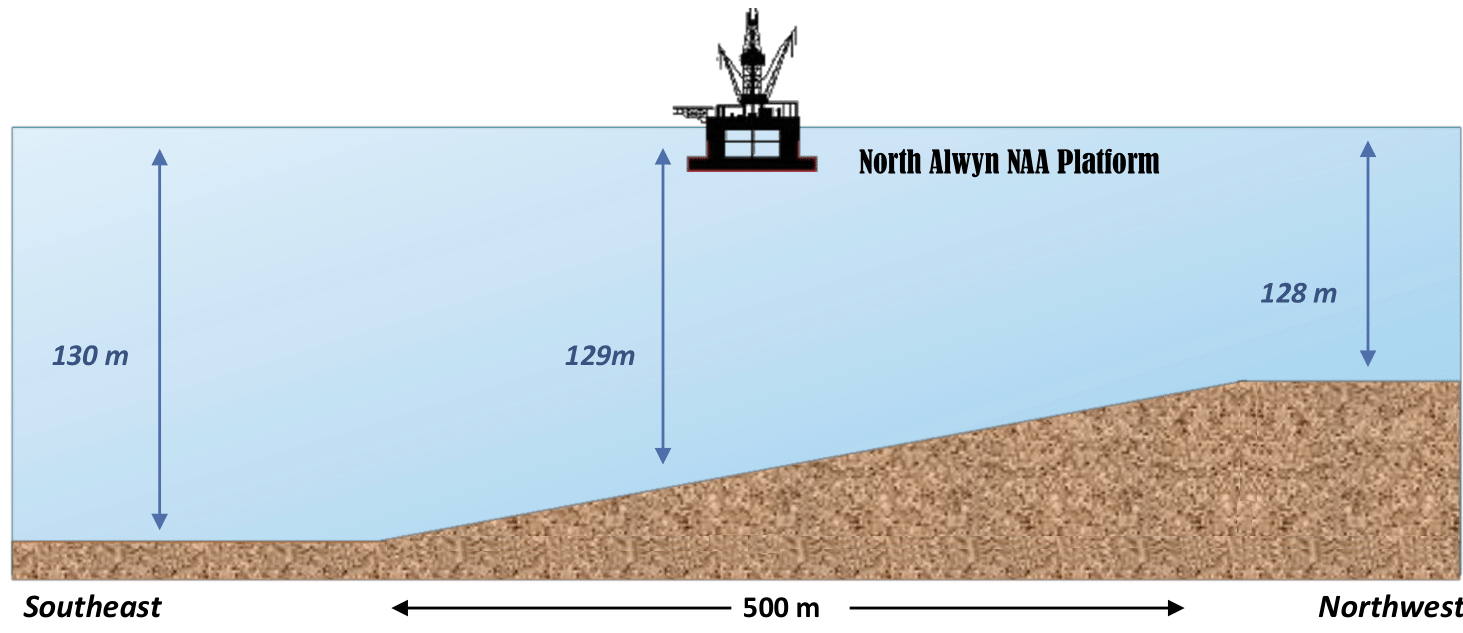}
\caption{\ryy{Approximate bathymetric features around the oil platform in the North Sea. The sketch is not up to scale.}}
\label{fig:sketch}
\end{figure}
\begin{eqnarray}
\nonumber
  \mu_{3} ( k_{p}h > \pi) \,\, &\approx & \,\, 3k_{1}\sigma (1- \nu \sqrt{2} + \nu^{2}) 
\\
&\equiv & 3k_{1}\sigma \cdot \mathfrak{B}(\nu)  \approx   \frac{\pi}{\sqrt{2}}\, \varepsilon \, \mathfrak{B}(\nu) \quad ,
\label{eq:tayfunmodel}
\end{eqnarray}
where $H_{s} \px{ = \pi \varepsilon / \sqrt{2} k_{p}}$ \px{and $k_{p}$ is the peak wavenumber obtained from} the spectral mean wavenumber $k_{1}$ through $k_{\px{p}} \approx (\px{3/4}) k_{\px{1}}$ \citep{Mendes2021b} \rr{and $\nu$ is the spectral bandwidth \citep{Higgins1975}}. In deep water \xxg{(}$k_{p}h \geqslant 5$\xxg{)}\ryy{,} \jfm{figure} \ref{fig:mu3steep}\jfm{a} shows that \ryy{the skewness is almost independent of the bandwidth, as expected from} eq.~(\ref{eq:tayfunmodel}). On the other hand, as the depth decreases to intermediate waters the ratio \rr{$\mu_{3}/\varepsilon$} significantly increases \ryy{and \rr{tends to} strongly depend on bandwidth}. \pk{To account for this f}inite depth effect\ryy{,} \rp{we rewrite} eq.~(\ref{eq:tayfunmodel}) \rp{according to} eq. 11 of \citet{Tayfun2020}:
\begin{equation}
\vvg{  \mu_{3} \, \pk{\approx} \, \frac{\pi \varepsilon}{\sqrt{2}}\, \pk{\mathfrak{B}}(\nu) \Big(\tilde{\chi}_{0}  + \frac{\sqrt{\tilde{\chi}_{1}}}{2} \Big) \quad ,}
\label{eq:Gnew0}
\end{equation}
with notation $\tilde{\chi}_{i}$ from \citet{Mendes2021b}:
\begin{eqnarray}
\nonumber
\tilde{\chi}_{0} = \frac{ \left[   4   \left(  1 + \frac{2k_{p}h}{\sinh{(2k_{p}h)}}  \right) - 2  \right]  }{ \left(  1 + \frac{2k_{p}h}{\sinh{(2k_{p}h)}}  \right)^{2} \tanh{k_{p}h} - 4 k_{p}h }  \quad ,
\\
\frac{\sqrt{\tilde{\chi}_{1}}}{2} =   \frac{3 - \tanh^{2}{(k_{p}h)} }{ 2\tanh^{3}{(k_{p}h)}  }\quad . 
\label{eq:tayfunmodel2}
\end{eqnarray}
\qs{Although \citeauthor{Tayfun2020}'s model \rp{provides} a good fit \rp{of $\mu_{3}/\varepsilon$} for $k_{p}h > \ryy{3}$}, the sum $\tilde{\chi}_{0} + \sqrt{\tilde{\chi}_{1}} / 2$ \px{stays close to unity} for $k_{p}h \, \qs{\geqslant} \, 2$. \qs{Hence, the larger values of the ratio $\mu_{3}/\varepsilon$ for shallower water ($2 \leqslant k_{p}h \leqslant \pi$) must stem from a dependence of $\mathfrak{B}(\nu)$ with depth. We therefore seek a \ryy{generalization of eq.~(\ref{eq:Gnew0}) whereby we fit a} function $\mathfrak{B}(\nu , k_{p}h) = 1- \nu \sqrt{2} +f_{k_{p}h} \cdot \nu^{2}$ \qx{capable of providing a} smooth transition from $f_{k_{p}h \sim \qx{3}} \qx{\approx 3.5}$ \qx{in shallower depths (see \jfm{figure} \ref{fig:mu3steep}\jfm{a})} to the deep water value $f_{k_{p}h = \infty} \sim 1$ \qx{(see eq.~(\ref{eq:tayfunmodel}))}. \qx{Hence,} \ryy{implementing this fit into eq\rp{s}.~(\ref{eq:betalpha2},\ref{eq:betalpha22})} the vertical asymmetry accounting for depth-induced effects is of the type:}
\begin{figure}
\begin{subfigure}[t]{0.43\textwidth}
    \includegraphics[scale=0.52]{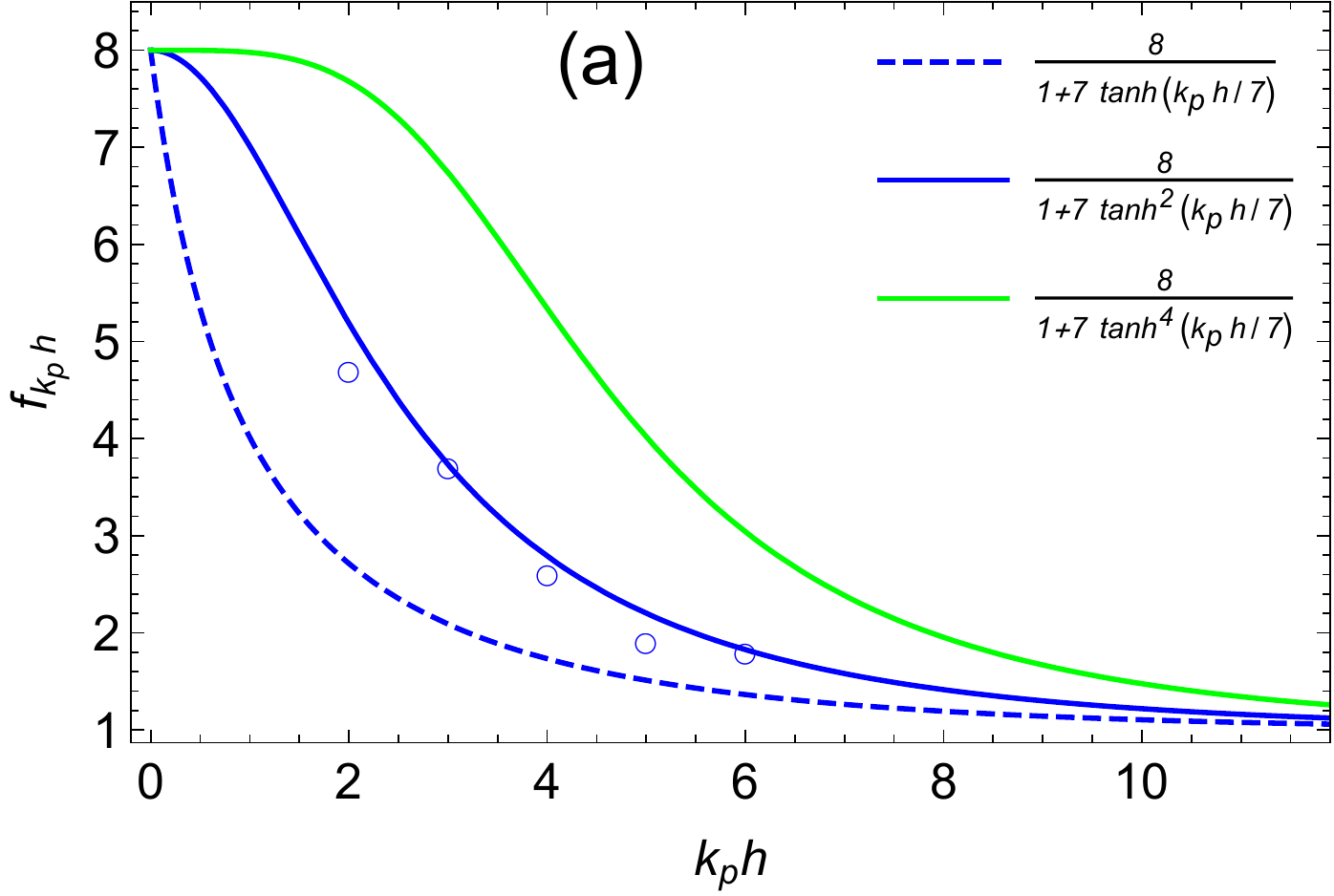}
\end{subfigure}

\begin{subfigure}[t]{0.45\textwidth}
    \includegraphics[scale=0.5]{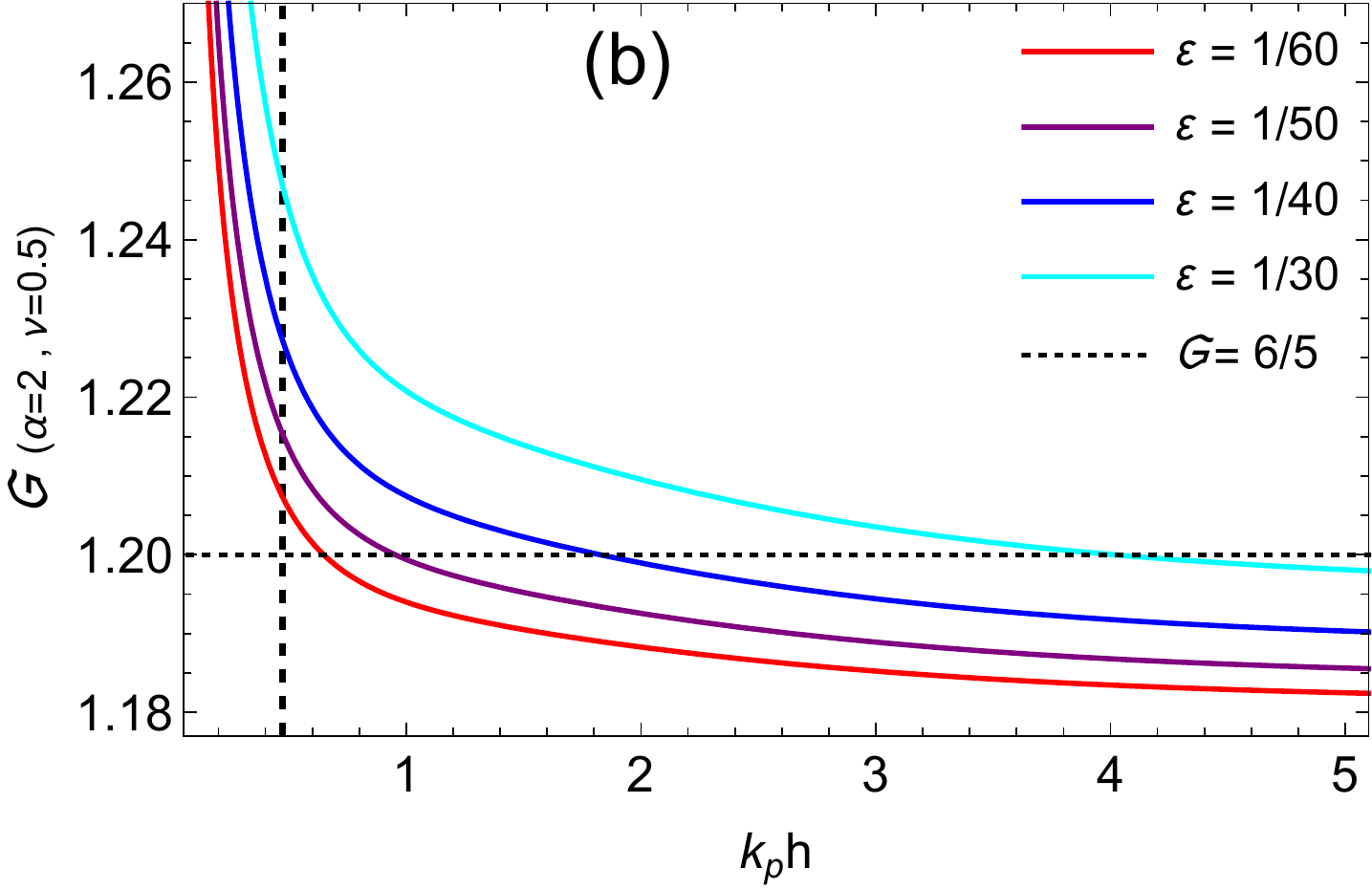}
\end{subfigure}
\caption{(a) Finite-depth functions $f_{k_{p}h}$ versus data (\qq{circles}) from \jfm{figure} \ref{fig:mu3steep}\jfm{a}. (b) Vertical asymmetry of broad-banded
rogue waves \qq{$(\nu = 0.5)$} as a function of water depth for different steepness, with the dotted line
depicting the empirical mean value $\mathfrak{S} = 6/5$ from \citet{Mendes2021a,Mendes2021b}. \qq{Dashed vertical line marks the limit of validity of second-order theory.}}
\label{fig:CD}
\end{figure}
\begin{figure*}
\minipage{0.3\textwidth}
  \includegraphics[width=5.2cm,height=5.8cm]{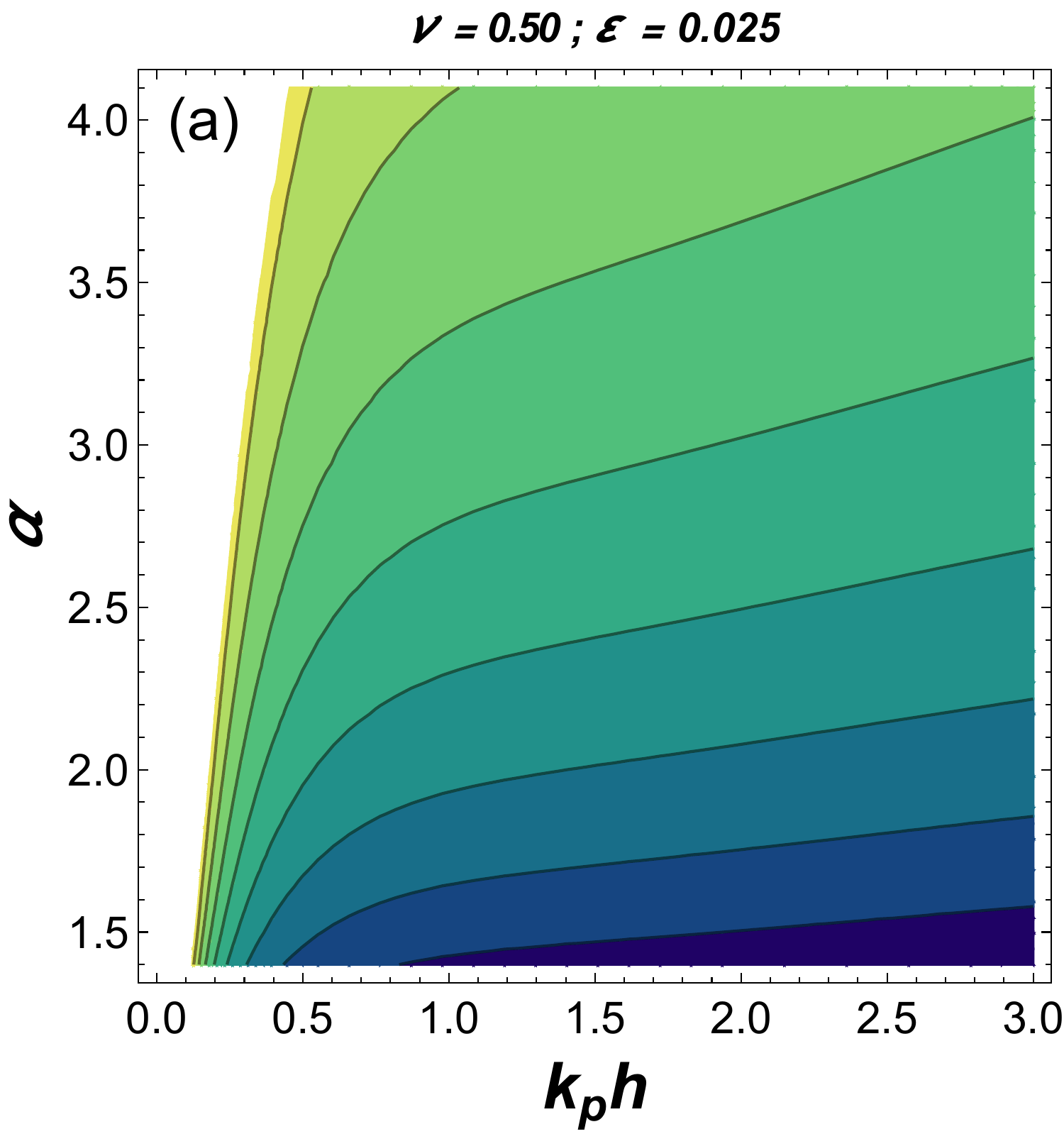}
\endminipage
\hfill
\hspace{-0.2cm}
\minipage{0.3\textwidth}
  \includegraphics[width=5.2cm,height=5.8cm]{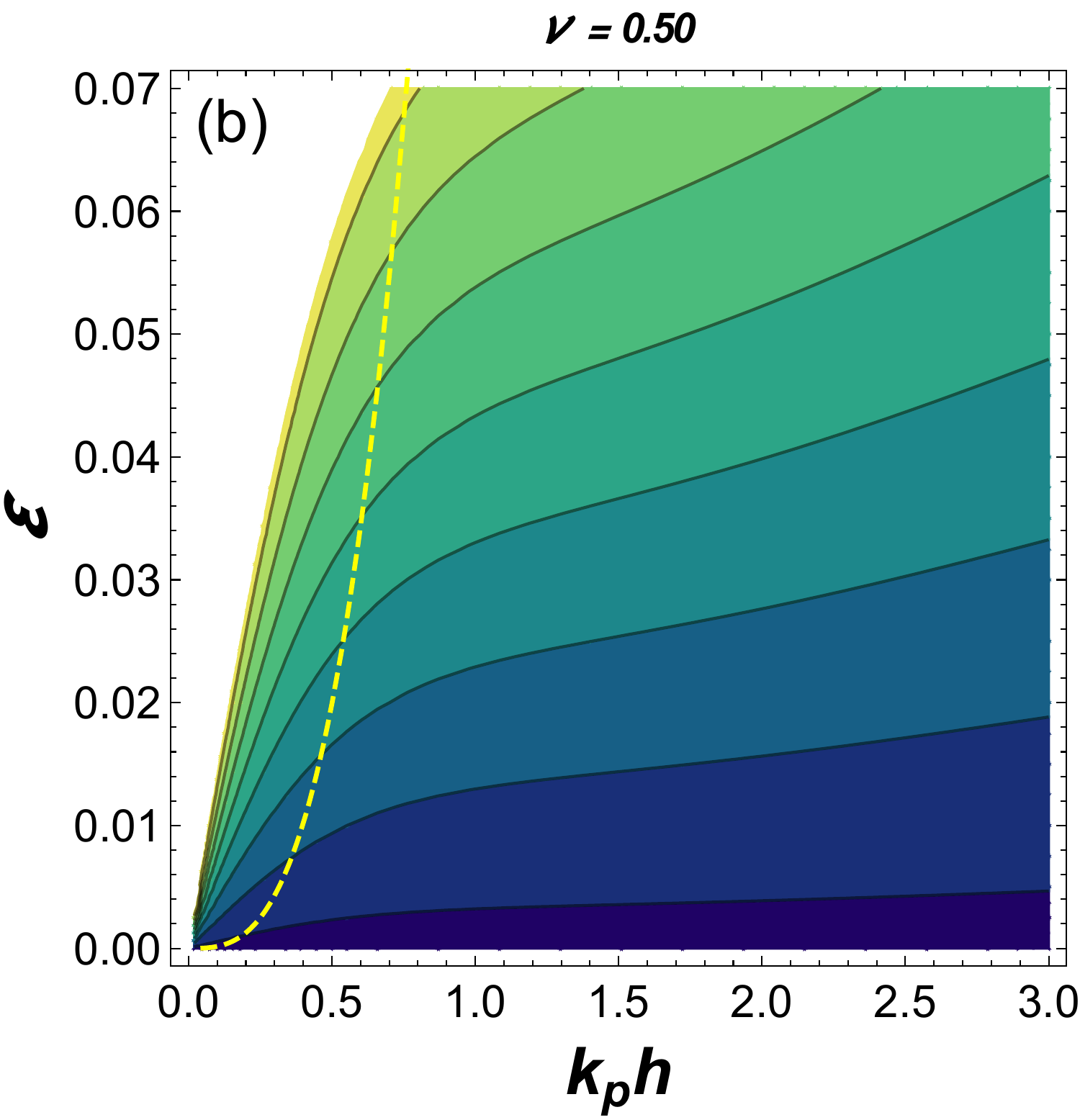}
\endminipage
\hfill
\minipage{0.35\textwidth}%
  \includegraphics[scale=0.43]{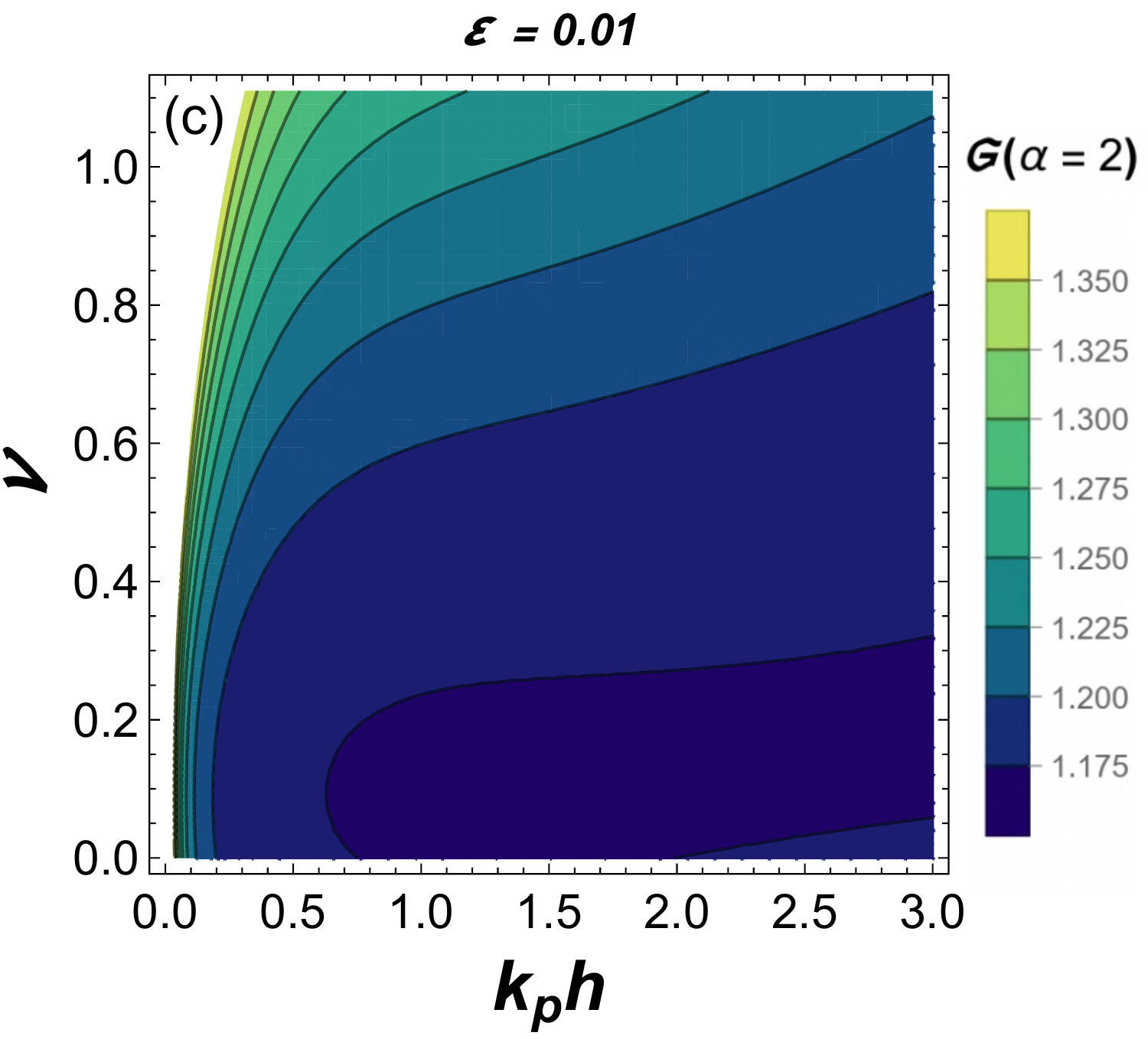}
\endminipage
\caption{Vertical asymmetry \pk{of large and} rogue waves as a function of water depth for different steepness\pk{, bandwidth and normalized height}. \qs{The} dashed line \qq{in panel b} represent\qs{s} the Ursell limit \qs{for} second-order theory.}
\label{fig:Contour}
\end{figure*}
\begin{equation}
\mathfrak{S}(\alpha = 2) \approx \frac{ (2+6\varepsilon_{\ast})(7+3\varepsilon_{\ast})  }{ 6 (2+3\varepsilon_{\ast})  } \quad ,
\label{eq:Gnew}
\end{equation}
\rr{where} $\qs{\varepsilon_{\ast}}$ is the effective steepness:
\begin{equation}
\varepsilon_{\ast} \approx \frac{\pi \varepsilon}{3\sqrt{2}}\,  \Big[1- \nu \pk{\sqrt{2}} +f_{\px{k_{p}h}} \cdot \nu^{2} \Big] \Big(\tilde{\chi}_{0}  + \frac{\sqrt{\tilde{\chi}_{1}}}{2} \Big) \quad .
\label{eq:effecsteep}
\end{equation}
\jfm{Figure} \ref{fig:mu3steep}\jfm{b} provides a contour plot \ryy{for the ratio $\mu_{3}/\varepsilon$ taking into account} the fitted model of \ryy{$f_{k_{p}h}$}. Here, $f_{k_{p}h}$ is a function of depth that can be obtained through the constraint \rr{$\mathfrak{S} \leqslant 2$} of eq~(\ref{eq:betalpha20}) applied to eq.~(\ref{eq:Gnew}):
\begin{equation}
\vvg{  \lim_{\substack{k_{p}h \rightarrow 0 }} \mathfrak{S}(\alpha = 2) \approx \lim_{\substack{k_{p}h \rightarrow 0  } } \frac{ (2+6\pp{\varepsilon_{\ast}})(7+3\pp{\varepsilon_{\ast}})  }{ 6 (2+3\pp{\varepsilon_{\ast}})  } \leqslant 2   \quad ,
} 
\label{eq:lim}
\end{equation}
\rp{thus leading to:}
\begin{equation}
  \vv{9}\pp{\varepsilon_{\ast}}^{2} + \vv{6} \pp{\varepsilon_{\ast}} - 5 \leqslant  0  \quad \therefore \quad \pp{\varepsilon_{\ast}} \leqslant \frac{\sqrt{6}-1}{3} \quad . 
\end{equation}
The function $\mathfrak{B}(\nu , k_{p}h)$ makes the exceedance probability of rogue waves weakly dependent on the bandwidth $\nu$  \citep{Higgins1975}. \rr{Very broad-banded} seas \rr{(}$\nu \, \rr{\geqslant} \, 1$\rr{)} \ryy{are very rare. For example, they} account for \ryy{only} 3\% of observed \qs{stormy} states \qs{in the North Sea} \citep{Mendes2021a}. \qs{These extreme sea conditions are typically short-lived and found for instance in hurricanes. Albeit bandwidths much larger than $\nu = 1$ can increase the vertical asymmetry \ryy{by} about 5-10\%, their lifespan \ryy{impacts} the weighted \rr{average of the} \qx{exceedance} probability \qx{of rogue waves} over a daily forecast \ryy{by only} $\sim 10\%$ \rr{because $\nu \sim 0.5$ over 97\% of all 30-min records}. Accordingly, we may set $\nu = 1$ as the realistic and effective maximum bandwidth to \qx{be considered for estimating} the \ryy{rogue wave} exceedance probability. Hence}, in the second-order limit \qs{we obtain}:
\begin{equation}
\vvg{ \lim_{k_{p}h \rightarrow 1/2} \,\, \frac{\pi \varepsilon}{3\sqrt{2}}\,  f_{\px{k_{p}h}} \Big(\tilde{\chi}_{0}  + \frac{\sqrt{\tilde{\chi}_{1}}}{2} \Big) < \frac{\sqrt{6}-1}{3} \quad , }
\label{eq:lim2}
\end{equation}
\rp{Consequently, broad-banded waves will not exceed the following depth correction:}
\begin{equation}
f_{\px{k_{p}h}} (\nu = 1) \lesssim \frac{18\sqrt{2}}{\pi} \approx 8 \quad .
\label{eq:lim3}
\end{equation}
Broad-banded waves have an \qs{effective steepness} of the order of $\varepsilon f_{k_{p}h} \pk{\nu^{2}}$\qx{.} \qx{S}ince finite-depth effects involve the ratio $\varepsilon / k_{p}h$ \ryy{which} it is directly related to $H_{s}/h$ \rr{and $f_{k_{p}h}$ grows quickly from deep to intermediate waters (see \jfm{figure} \ref{fig:mu3steep}\jfm{a})}\ryy{,} \qq{we expect $f_{k_{p}h}$ \rr{to be inversely proportional to the relative depth $k_{p}h$}.} \qq{In order to fulfill eqs.~(\ref{eq:lim}-\ref{eq:lim3}), a sigmoid function provid\rr{es}} \rr{a good} fit with continuous derivative for the North Sea data (see \jfm{figure} \ref{fig:CD}\jfm{a}):
\begin{equation}
f_{\px{k_{p}h}} \approx \frac{\qs{8}}{1+\qs{7} \, \tanh^{2}{( k_{p}h/7)}}  \quad , \quad \nu \leqslant \qs{1} \quad .
\label{eq:GG}
\end{equation}
\qq{Plugging eq.~(\ref{eq:GG}) into eq.~(\ref{eq:Gnew}) introduces an approximation for the vertical asymmetry covering the entire range of second-order theory for narrow and broad-banded irregular waves. In fact, \jfm{figure} \ref{fig:CD}\jfm{b}} \qq{shows that} the vertical asymmetry \px{is almost constant} \pp{for} \qq{typical values of} \rr{mean} steepness \qq{($\varepsilon \qs{\ll 1/10}$)} in intermediate and deep waters \px{($k_{p}h \geqslant \pk{\pi /10}$)}. \px{Conversely,} sharp increases in \px{the} \rr{mean} steepness will \pk{induce} a few percent increase in the vertical asymmetry \px{in the same regimes (}$k_{p}h \geqslant  \pk{\pi /10}$\px{)}. \qq{The contour plot in \jfm{figure} \ref{fig:Contour}\jfm{b} provides a full description of the variations in asymmetry with depth and steepness. Furthermore,} \jfm{figure} \ref{fig:Contour}\jfm{a} \qx{shows that in shallow depths the vertical asymmetry strongly depends on $k_{p}h$ while in deep water it tends to saturate}. \jfm{Figure} \ref{fig:Contour}\jfm{c} \qq{also \ryy{illustrates} the role of} bandwidth \qq{in increasing the asymmetry, albeit sharp changes are restricted to sufficiently broad spectra ($\nu > 0.8$)}.  \pp{Thus}, \rr{\rp{the} analysis \rp{of}  field data \rp{from the North Sea} shows that as long as} the steepness \px{in intermediate water ($k_{p}h \, \rr{>} \, \pi /10$) is \rr{small} ($\varepsilon \, \rr{<} \, 1/10$)} \pk{or the spectrum \rr{narrow} $(\nu \, \rr{<} \, 1/2)$}, the vertical asymmetry \rp{stays close to} $\mathfrak{S}= \rxx{1.2}$. \rr{W}e find th\rp{is} approximation for the vertical asymmetry to be \rr{still} applicable to the experiments in \citet{Trulsen2020} \rr{with steeper slope, as shown in} \jfm{appendix} \ref{sec:App}.

\vvg{\px{Moreover}, \px{the special case of} narrow-banded \px{(}$\nu = 0$\px{)} \px{linear} waves \px{(}$\varepsilon \ll 1/10${) in deep water leads to \pk{$\varepsilon_{\ast} \rightarrow 0$, thus reaching} \rr{the} lower bound \rr{of}} the \pk{asymmetry} $\mathfrak{S} \, \qq{=} \, 7/6$ \px{\pk{for rogue waves}. This} suggests that in intermediate waters \px{\qq{narrowing} the bandwidth} \px{from} $\nu = 0.3$ \px{to $\nu = 0$} will have little \px{impact on the} amplification of rogue wave statistics \px{due to the \pk{negligible change} in vertical asymmetry, whereas \pk{in shallow water} increasing \qs{the} bandwidth above $\nu = 0.5$ will significant}\rr{ly boost rogue wave occurrence}. From the point of view of the theory \px{in} \citet{Mendes2021b}, the \xxg{asymmetry approximation} \px{of eqs.~(\ref{eq:Gnew},\ref{eq:GG})} explain\px{s} why narrow-banded models \citep{Adcock2021c} are successful in predicting rogue wave statistics travelling past a step in a broad-banded irregular wave background \pk{in intermediate water. \qq{P}rovided the\qq{re} is no wave breaking $(H_{s}/h \ll 1)$, the bandwidth effect will play a role in amplifying statistics in shallower depths because \qq{of the contributuin of the term} $f_{k_{p}h}\nu^{2}$\qq{, as e}xperimental\qq{ly} demonstrated in \citet{Doeleman2021}}.}

\section{UPPER BOUND FOR KURTOSIS}\label{sec:KurtApp}

\rxx{The excess kurtosis \ryy{has been used} in the past two decades as a proxy for \rr{how} rough nonlinear seas \rr{increase the occurrence and intensity of rogue waves}\ryy{. Therefore, in this section we extend our results of \jfm{section} \ref{sec:kurtosis} to estimate the maximum kurtosis atop a\rr{ny} shoal} \rr{in} the ocean \citep{Janssen2009,Janssen2017}.} The assess\px{ment of} maximum expected waves over a \vv{specific return} time \rxx{at a fixed location} \px{is crucial} \rxx{f}or \vv{naval} design. Typically, \vvg{ocean} structure\vvg{s and vessels} must be designed to sustain \vvg{expected} maximum extreme wave\pp{s} over \vv{their} life\xxg{s}pan \citep{Borgman1973,El-Shaarawi1986}. In order to do so, we shall evaluate maxima for the parameters $\mathfrak{S}$ \ryy{and} $\Gamma$. \qs{E}qs.~(\ref{eq:Gnew}) \pk{and (}\ref{eq:GG}) \qs{provide} the upper bound \qs{for the} vertical asymmetry \qs{of} rogue wave\px{s} \ryy{in the limit of wave breaking}: $\mathfrak{S}_{\infty}(k_{p}h = \infty) \approx 1.387$ \px{in deep water \qs{and}} $\mathfrak{S}_{\infty}(k_{p}h = 0) \approx 1.668$ \px{in shallow water}. Since the \ryy{$\Gamma$ correction is also} limited \qs{by} wave breaking, one finds the bound $\px{\Gamma_{\infty} - 1} \lesssim 1/12$ \pk{due to} eq. (3.17) \pk{of} \citet{Mendes2021b}. \pk{Hence,} we may approximate:
\begin{eqnarray}
1 - \frac{1}{\mathfrak{S}^{2}_{\infty} \Gamma_{\infty}} \, \rp{\lesssim} \, 8 (\Gamma_{\infty} - 1) \quad .
\label{eq:Gapprox}
\end{eqnarray}
\px{\qx{Approaching the value} $\Gamma_{\infty}$ atop the shoal \ryy{(region III of \jfm{figure} \ref{fig:shoal})}, the contribution of the skewness \rr{to} the amplification of wave statistics \rr{near the breaking regime} increases such that the relationship between kurtosis and skewness leading to eq.~(\ref{eq:Mori2X}) \qx{is modified and} now} \rr{empirically reduces to} $\mu_{4} \, \pp{\approx} \, \mu_{3}^{2}$ \citep{Ma2015}. \ryy{Plugging this relationship into eq.~(\ref{eq:Mori1X}) and comparing it with eqs.~(\ref{eq:bath},\ref{eq:Gapprox}), we obtain:}
\begin{equation}
e^{\vvg{16}\alpha^{2}  (\Gamma_{\qs{\infty}} - 1)}\, \vvg{\geqslant} \, 1 +  \alpha^{2} \left( \alpha^{2} - 1  \right) \mu_{4} \quad .
\end{equation}
\rp{At} $\alpha = 2$\rp{, the evaluation of} \ryy{the excess kurtosis \rp{lies} at the region of stability of the Gram-Charlier series \rp{and}} \px{we are able to compute the} \vvg{upper} bound for the \vvg{excess} kurtosis \jkv{\ryy{in the case} of a pre-shoal Gaussian statistics} (see \jfm{figure} \ref{fig:KurtslopeGamma}):
\begin{figure}
\centering
  \includegraphics[scale=0.66]{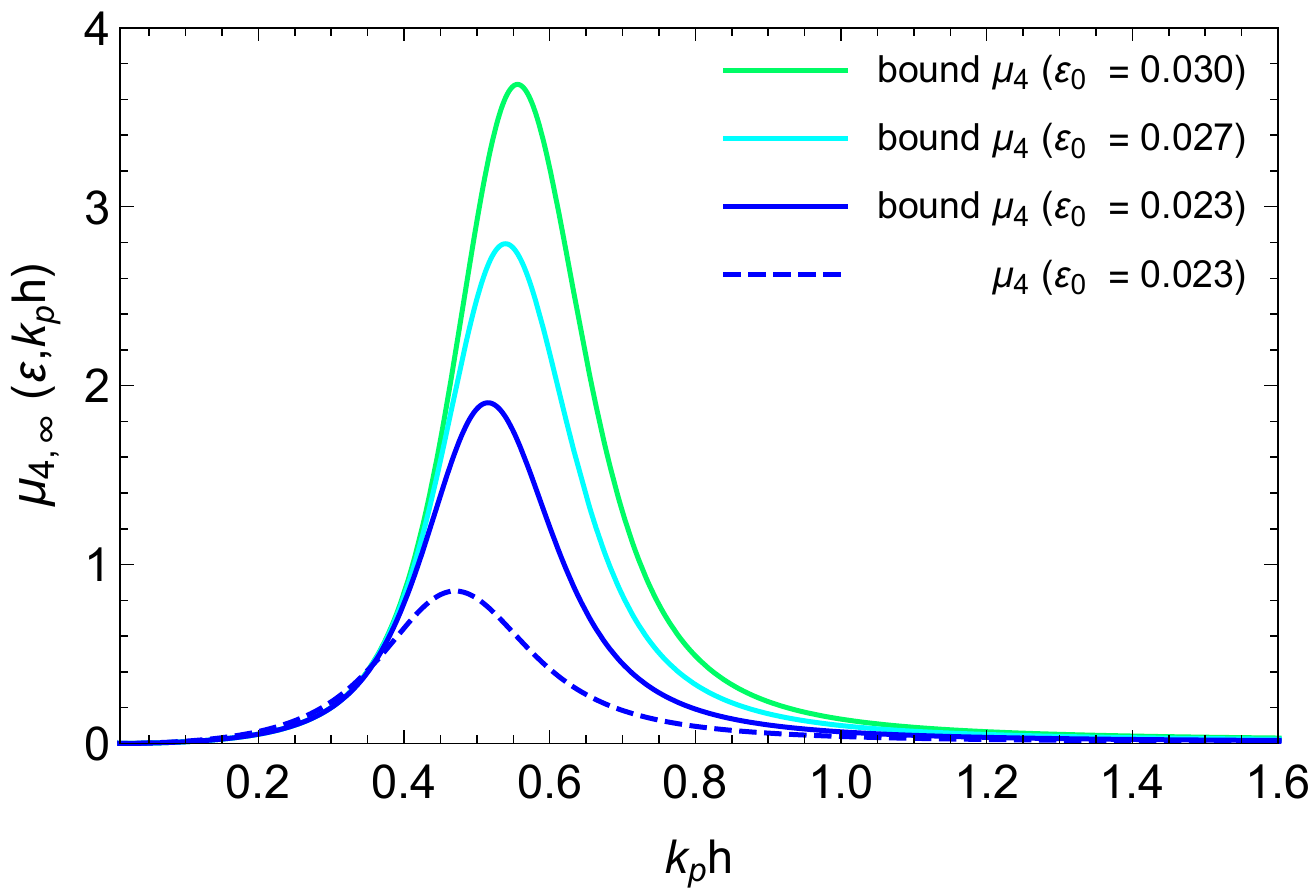}
\caption{Upper bound on kurtosis from eq.~(\ref{eq:kurtN2}) \vv{\pk{for $\nu=0.5$ and different} \px{pre-shoal} steepness \px{$\varepsilon_{0}$} \px{subject to linear shoaling}. \px{The dashed curve represents the computed kurtosis in \jfm{figure} \ref{fig:modelkurt1}\jfm{a}\pk{, representative of Run 1 of \citet{Trulsen2020}}}}. 
}
\label{fig:KurtslopeGamma}
\end{figure}
\begin{equation}
\mu_{4 \pk{\, , \, \infty}}  \approx   \frac{1}{12} \left[ e^{ 64 (\Gamma_{\qs{\infty}} - 1)} - 1  \right]   \quad \qs{,}
\label{eq:kurtN2}
\end{equation}
\qs{where $\Gamma_{\infty}$ varies with water depth.}
\vvg{\px{According to} eq.~(\ref{eq:kurtN2})}, \pk{\rr{typical seas with }steep and highly asymmetrical broad-banded waves} \px{lead} \qs{to an upper bound \qx{for} \ryy{the excess} kurtosis of the order of $\mu_{4 \, , \, \infty} \sim 4$} \rr{in intermediate water}, see \jfm{figure} \ref{fig:KurtslopeGamma}. We already described that \rr{the maximum value of} $\Gamma$ \rr{is located} around $k_{p}h \approx 0.5$ in \citet{Mendes2021b} \ryy{and eq.~(\ref{eq:Mori3X}) has been validated} in \jfm{figure} \ref{fig:modelkurt1}. \rp{Therefore,} the peak \rr{in} excess kurtosis will also \rr{be located} in this region. Experiment\rp{s conducted} in \citet{Benoit2023} \rp{found} the \rr{peak in excess} kurtosis \rr{in the \rp{same}} region \rr{$k_{p}h \approx 0.5$}.

\section{CONCLUSIONS}

In this work we have extended the \px{framework} in \citet{Mendes2021b} to \px{an effective theory for} \qq{the evolution of} excess kurtosis \qq{of the surface elevation} over a shoal of finite and constant \vvg{steep} slope. \px{W}e find quantitative agreement \ryy{with} experiments in \citet{Trulsen2020} \ryy{regarding the magnitude of the kurtosis increase} \px{during and atop the shoal}. \ryy{While} the groundwork of \citet{Marthinsen1992} computes \px{the excess kurtosis} directly from the solution $\zeta(x,t)$, \ryy{our model unravels} the kurtosis \ryy{dependence} on the inhomogeneities o\px{f} the energy density \pk{over a shoal}. \rxx{Our formulation outperform\ryy{s \ryy{the} conventional method \ryy{of \citet{Marthinsen1992}}} for the computation of kurtosis of the bound wave contribution. \ryy{In addition,} our effective theory is capable of describing changes of the kurtosis magnitude \ryy{over} arbitrary slopes \rr{provided} reflection \rr{can be neglected}.} \vvg{\rp{A} \px{computation} of the kurtosis \px{from the probability density of $\zeta(x,t)$} through the non-homogeneous framework \px{will be pursued} in a future work with \px{an} analytical non-uniform distribution of random phases.}

\px{F}urther\px{more,} \vvg{\px{w}e have  obtained a\vv{n} approximation for the vertical asymmetry in finite depth as a function of both steepness and bandwidth\px{. This approximation} \ryy{extends} the seminal work of \citet{Tayfun2006} for \qq{the skewness of the surface elevation} \ryy{to  broad-banded intermediate water waves while} recover\ryy{ing its original formulation for} narrow-banded deep water waves. Building on this new approximation, we have \px{demonstrated} that the vertical asymmetry \px{varies slowly} over a shoal \px{in both deep and intermediate waters}. \pp{Moreover}, \ryy{based on th\rp{is}} \rr{rise in vertical asymmetry} we were able to compute \rr{an upper bound for the} excess kurtosis driven by shoaling.}

\section{ACKNOWLEDGMENTS}

S.M and J.K. were supported by the Swiss National Science Foundation under grant 200020-175697. We thank Maura Brunetti and Alexis Gomel for fruitful discussions.

\appendix

\section{Computation of \rr{Irregular} Bound Wave Kurtosis}\label{sec:AppB}

\rxx{\rr{T}he contribution \rr{of} bound waves to \rr{the excess} kurtosis \rr{of the surface elevation is} given by \citet{Mori1998} in the regular wave approximation:}
\begin{equation}
\rxx{\mu_{4}  = 3 \left\{ \frac{  1 + (ka)^{2} ( 6D_{1}^{2} + 6 D_{2}^{2} + 8 D_{1}D_{2})   \Big] }{ \Big[ 1 + (ka)^{2} ( D_{1}^{2} +  D_{2}^{2})   \Big]^{2}  } - 1 \right\} \,\, , }
\label{eq:Morii}
\end{equation}
\rr{where $D_{1}$ and $D_{2}$ are relative water depth coefficients from the surface elevation:}
\begin{equation}
\rxx{D_{1} = \frac{1}{\tanh{kh}} \,\, ; \,\, D_{2} = D_{1} \Big( 1 + \frac{3}{ 2 \sinh^{2}{kh}}   \Big) \, .}
\end{equation}
\rr{To leading order in steepness, we may approximate the excess kurtosis as:}
\begin{eqnarray}
\nonumber
\rr{\mu_{4}} &\rr{\approx}& \rr{ 3 \Big\{  \Big[ 1 +  (ka)^{2} ( 6D_{1}^{2} + 6 D_{2}^{2} + 8 D_{1}D_{2})   \Big] }
\\
\nonumber
&\times & \Big[ 1 - 2 (ka)^{2} ( D_{1}^{2} +  D_{2}^{2})   \Big]  - 1       \Big\}  \, ,
\\
\nonumber
&\rr{\approx}& \rr{ 3 \left\{  \Big[ 1 +  (ka)^{2} ( 4D_{1}^{2} + 4 D_{2}^{2} + 8 D_{1}D_{2})   \Big]   - 1       \right\}  \, ,}
\\
\nonumber
&\approx& 3  (ka)^{2} ( 4D_{1}^{2} + 4 D_{2}^{2} + 8 D_{1}D_{2})
\\
&\approx & 12 (ka)^{2} (D_{1}+D_{2})^{2}
 \quad ,
\end{eqnarray}
\rxx{However, we \rr{shall extend} eq.~(\ref{eq:Morii}) for irregular wave\rr{s, computing the equivalent irregular mean wave} steepness \rr{and relative depth}. \rr{We may use $ka \rightarrow k_{p}H_{s}/2\sqrt{2}$ as pointed out in \citet{Trulsen2020} for irregular waves, consequently we find $ka \rightarrow (\pi / 4) \varepsilon$ as in \citet{Mendes2021b}. Hence, we may write the excess kurtosis as a function of $\varepsilon$ up to second order in steepness:}}
\begin{eqnarray}
\rxx{\mu_{4}} \, \rxx{\approx} \, \rxx{\frac{\rr{3}\pi^{2} }{\rr{4}} \cdot \varepsilon^{2} (D_{1}+D_{2})^{2}  \quad . }
\end{eqnarray}
\rxx{Moreover, the depth $kh$ has to be converted to \rp{its} peak wavenumber equivalent $k_{p}h$. Hence, since observations in the ocean feature $1.1 \leqslant \lambda_{p} / \lambda_{1/3} \leqslant 1.2 $ \citep{Figueras2010}, we may use the transformation from regular wave to irregular wave $kh \rightarrow 1.2 k_{p}h$ \rp{to compute $(D_{1},D_{2})$ correctly.}} \rr{As a remark, the above expression} differs little from formulations such as of \citet{Marthinsen1992} \ryy{and others} \rp{as} reviewed in \citet{Tayfun2020}.

\section{Slope Effect on Vertical Asymmetry}\label{sec:App}

\begin{figure}
\minipage{0.45\textwidth}
    \includegraphics[scale=0.46]{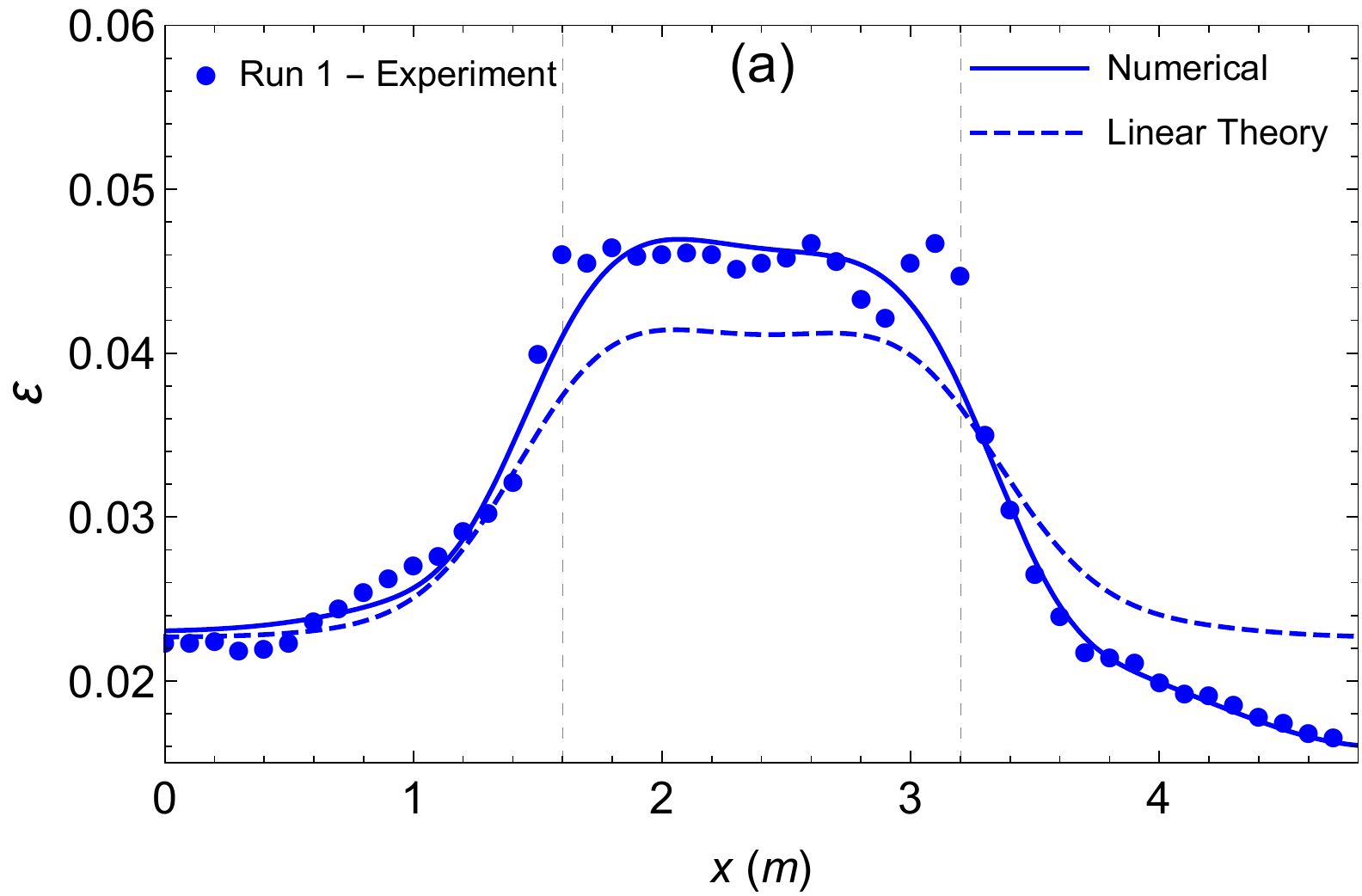}
\endminipage
\hfill
\minipage{0.48\textwidth}
    \includegraphics[scale=0.46]{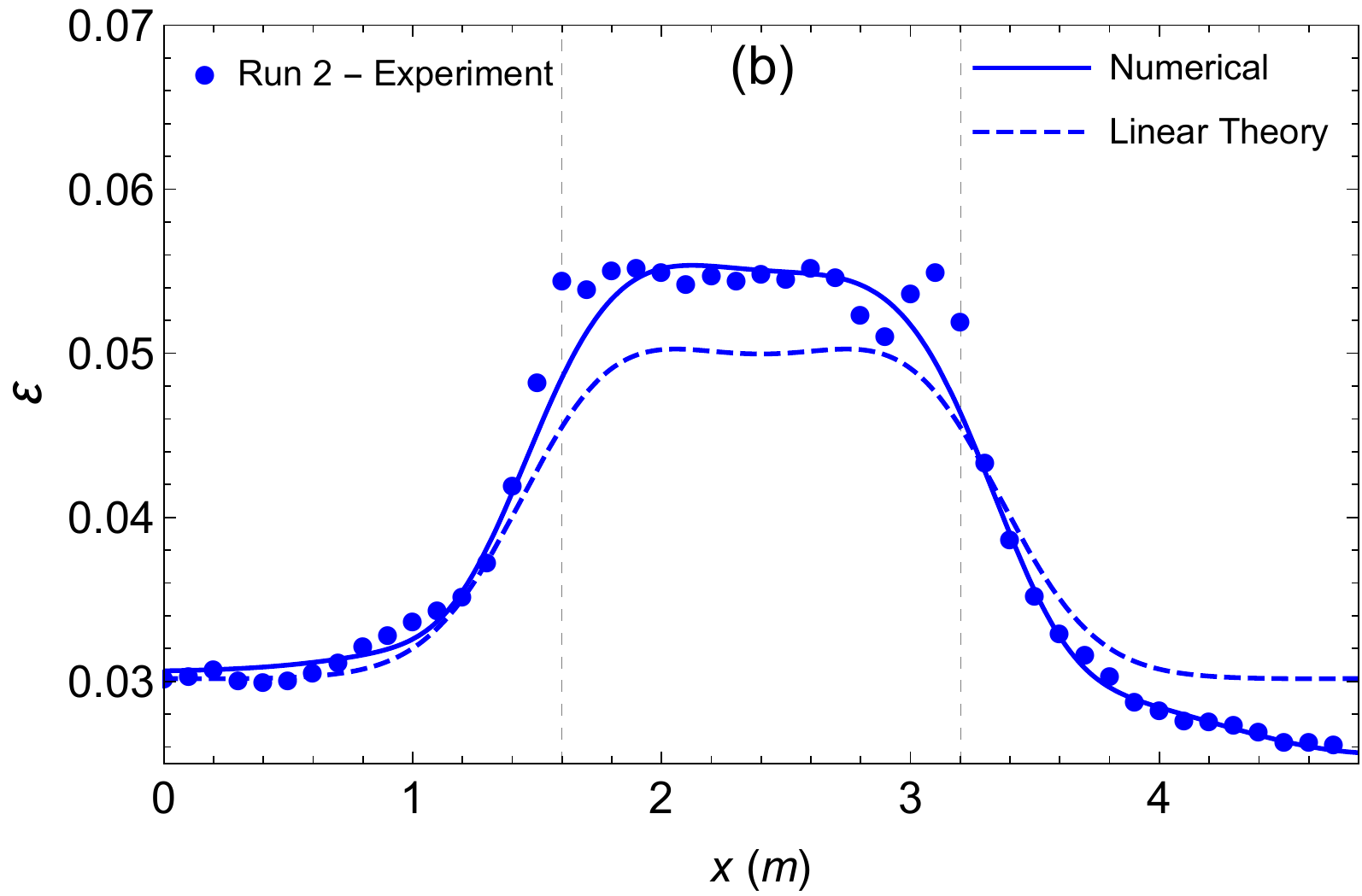}
\endminipage
\caption{\rxx{Theoretical \rp{evolution of} steepness measured against observations (dots) in \citet{Raustol2014} and its numerical fit thereof for (a) Run 1 and (b) Run 2 of the experiments in \citet{Trulsen2020}.}}
\label{fig:steepshoaling}
\end{figure}
\rp{In this section we assess how the vertical asymmetry of irregular rogue waves is affected by an arbitrary slope.}
\rxx{Let us denote the final steepness atop \rr{the} shoal as $\varepsilon_{\rr{f}}$ and the initial one as $\varepsilon_{0}$. If linear \rp{waves travel over a} shoal, then we may \ryy{define the amplification ratio of the steepness (also known as shoaling coefficient)}:}
\begin{equation}
\rxx{ K_{\varepsilon \, , \, \textrm{L}} := \frac{\varepsilon_{\rr{f}}}{\varepsilon_{0}} \approx \frac{1}{\tanh{(1.2 k_{p}h)}} \left[ \frac{2\cosh^{2}{(1.2 k_{p}h)}}{2.4 k_{p}h + \sinh{(2.4 k_{p}h})} \right]^{1/2}  , }
\label{eq:varepsilons1}
\end{equation}
\rxx{where we have converted the regular wave formula \citep{Holthuijsen2007} to irregular waves. Indeed, except for a few percent, the shoaling coefficient of the \rr{(irregular) significant} wave height is a good approximation for the regular wave \rr{counterpart} \citep{Goda1975,Goda2010}. If nonlinear wave shoaling is dominant, then $K_{\varepsilon \, , \, \rr{\textrm{NL}}}$ \rr{depends on the slope of the shoal} $\nabla h$\rr{, and} we denote the ratio $K_{\varepsilon \, , \, \rr{\textrm{NL}}} / K_{\varepsilon \, , \, \rr{\textrm{L}}} = \mathcal{F}_{\nabla h}$ \citep{Eagleson1956,Walker1983,Srineash2018}.} \rr{Performing a Taylor expansion in eq.~(\ref{eq:Gnew}) up to first order in $\varepsilon_{\ast}$, the vertical asymmetry of small wave amplitudes can be written as:}
\begin{equation}
\rxx{  \mathfrak{S}(\alpha = 2) \approx \frac{7}{6} \big(1 + 2 \varepsilon_{\ast} \big)  \quad . }
\end{equation}
\rr{The typical sea representative of \citet{Trulsen2020} experiments is broad-banded $(\nu \sim 0.5)$ and in intermediate water ($k_{p}h \sim 1$). Recalling eqs.~(\ref{eq:effecsteep},\ref{eq:GG}), this leads to $\mathfrak{B}(\nu) \sim 2$ and $\tilde{\chi}_{0} + \sqrt{\tilde{\chi}_{1}} / 2 \sim 1$. Therefore we may approximate $\varepsilon_{\ast} \approx (\pi \sqrt{2}/3) \, \varepsilon$. Consequently, the ratio between the vertical asymmetry of identical sea states of waves travelling over a shoal of different slopes is approximately described by the formula:}
\begin{equation}
\frac{ \mathfrak{S}_{|\nabla h| } }{ \mathfrak{S}  } \approx  \frac{  \left(1 + \frac{2 \sqrt{2}\pi}{3} \varepsilon \cdot \mathcal{F}_{\nabla h} \right)  }{  \left(1 + \frac{2 \sqrt{2}\pi}{3} \varepsilon \right) } 
\approx  1 + \frac{2 \sqrt{2}\pi}{3} \varepsilon \left(  \mathcal{F}_{\nabla h} - 1  \right)  \, . 
\end{equation}
\rxx{Even for relatively steep shoals ($|\nabla h| \approx 1/4$) as in the case of \citet{Trulsen2020} the correction accounting for slope is \rp{small}, with $\mathcal{F}_{\nabla h} \approx 1.15$ in this case (see \jfm{figure} \ref{fig:steepshoaling}\jfm{a,b}). \rr{In fact, \citet{Srineash2018} demonstrated experimentally that $\mathcal{F}_{\nabla h} - 1$ stays in the range of $0.1-0.2$ for steep slopes.}}
\rxx{Since the steepness in the experiments of \citet{Trulsen2020} \rp{atop the shoal} do\rp{es} not exceed $\varepsilon = 0.06$, the slope correction to the vertical asymmetry derived with the help of field data from the North Sea \rp{stay below} $(\pi \sqrt{2}/\rr{9}) \times 100\% \times 0.06 = 3\% $. Thus, eq.~(\ref{eq:Gnew}) is applicable to the analysis in \jfm{section} \ref{sec:kurtosis}, and the approximation $\mathfrak{S}\approx 1.2$ \rp{is applicable in} the conditions of the experiments in \citet{Trulsen2020}.}

\bibliography{Maintext}

\begin{thebibliography}{56}%
\makeatletter
\providecommand \@ifxundefined [1]{%
 \@ifx{#1\undefined}
}%
\providecommand \@ifnum [1]{%
 \ifnum #1\expandafter \@firstoftwo
 \else \expandafter \@secondoftwo
 \fi
}%
\providecommand \@ifx [1]{%
 \ifx #1\expandafter \@firstoftwo
 \else \expandafter \@secondoftwo
 \fi
}%
\providecommand \natexlab [1]{#1}%
\providecommand \enquote  [1]{``#1''}%
\providecommand \bibnamefont  [1]{#1}%
\providecommand \bibfnamefont [1]{#1}%
\providecommand \citenamefont [1]{#1}%
\providecommand \href@noop [0]{\@secondoftwo}%
\providecommand \href [0]{\begingroup \@sanitize@url \@href}%
\providecommand \@href[1]{\@@startlink{#1}\@@href}%
\providecommand \@@href[1]{\endgroup#1\@@endlink}%
\providecommand \@sanitize@url [0]{\catcode `\\12\catcode `\$12\catcode
  `\&12\catcode `\#12\catcode `\^12\catcode `\_12\catcode `\%12\relax}%
\providecommand \@@startlink[1]{}%
\providecommand \@@endlink[0]{}%
\providecommand \url  [0]{\begingroup\@sanitize@url \@url }%
\providecommand \@url [1]{\endgroup\@href {#1}{\urlprefix }}%
\providecommand \urlprefix  [0]{URL }%
\providecommand \Eprint [0]{\href }%
\providecommand \doibase [0]{https://doi.org/}%
\providecommand \selectlanguage [0]{\@gobble}%
\providecommand \bibinfo  [0]{\@secondoftwo}%
\providecommand \bibfield  [0]{\@secondoftwo}%
\providecommand \translation [1]{[#1]}%
\providecommand \BibitemOpen [0]{}%
\providecommand \bibitemStop [0]{}%
\providecommand \bibitemNoStop [0]{.\EOS\space}%
\providecommand \EOS [0]{\spacefactor3000\relax}%
\providecommand \BibitemShut  [1]{\csname bibitem#1\endcsname}%
\let\auto@bib@innerbib\@empty
\bibitem [{\citenamefont {Clauss}(2002)}]{Clauss2002}%
  \BibitemOpen
  \bibfield  {author} {\bibinfo {author} {\bibfnamefont {G.}~\bibnamefont
  {Clauss}},\ }\bibfield  {title} {\bibinfo {title} {Dramas of the sea:
  episodic waves and their impact on offshore structures},\ }\href@noop {}
  {\bibfield  {journal} {\bibinfo  {journal} {App. Ocean Res.}\ }\textbf
  {\bibinfo {volume} {24}},\ \bibinfo {pages} {147} (\bibinfo {year}
  {2002})}\BibitemShut {NoStop}%
\bibitem [{\citenamefont {Toffoli}\ \emph {et~al.}(2015)\citenamefont
  {Toffoli}, \citenamefont {Waseda}, \citenamefont {Houtani}, \citenamefont
  {Cavaleri}, \citenamefont {Greaves},\ and\ \citenamefont
  {Onorato}}]{Toffoli2015}%
  \BibitemOpen
  \bibfield  {author} {\bibinfo {author} {\bibfnamefont {A.}~\bibnamefont
  {Toffoli}}, \bibinfo {author} {\bibfnamefont {T.}~\bibnamefont {Waseda}},
  \bibinfo {author} {\bibfnamefont {H.}~\bibnamefont {Houtani}}, \bibinfo
  {author} {\bibfnamefont {L.}~\bibnamefont {Cavaleri}}, \bibinfo {author}
  {\bibfnamefont {D.}~\bibnamefont {Greaves}},\ and\ \bibinfo {author}
  {\bibfnamefont {M.}~\bibnamefont {Onorato}},\ }\bibfield  {title} {\bibinfo
  {title} {Rogue waves in opposing currents: an experimental study on
  deterministic and stochastic wave trains},\ }\href@noop {} {\bibfield
  {journal} {\bibinfo  {journal} {Journal of Fluid Mechanics}\ }\textbf
  {\bibinfo {volume} {769}},\ \bibinfo {pages} {277–297} (\bibinfo {year}
  {2015})}\BibitemShut {NoStop}%
\bibitem [{\citenamefont {Haver}(2004)}]{Haver2004}%
  \BibitemOpen
  \bibfield  {author} {\bibinfo {author} {\bibfnamefont {S.}~\bibnamefont
  {Haver}},\ }\bibfield  {title} {\bibinfo {title} {A possible freak wave event
  measured at the draupner jacket january 1 1995},\ }\href@noop {} {\bibfield
  {journal} {\bibinfo  {journal} {Proc. Rogue Waves 20-22 October IFREMER}\ }
  (\bibinfo {year} {2004})}\BibitemShut {NoStop}%
\bibitem [{\citenamefont {Akhmediev}\ \emph {et~al.}(2009)\citenamefont
  {Akhmediev}, \citenamefont {Ankiewicz},\ and\ \citenamefont
  {Taki}}]{Akhmediev2009x}%
  \BibitemOpen
  \bibfield  {author} {\bibinfo {author} {\bibfnamefont {N.}~\bibnamefont
  {Akhmediev}}, \bibinfo {author} {\bibfnamefont {A.}~\bibnamefont
  {Ankiewicz}},\ and\ \bibinfo {author} {\bibfnamefont {M.}~\bibnamefont
  {Taki}},\ }\bibfield  {title} {\bibinfo {title} {Waves that appear from
  nowhere and disappear without a trace},\ }\href@noop {} {\bibfield  {journal}
  {\bibinfo  {journal} {Phys. Lett. A}\ }\textbf {\bibinfo {volume} {373}},\
  \bibinfo {pages} {675} (\bibinfo {year} {2009})}\BibitemShut {NoStop}%
\bibitem [{\citenamefont {Rice}(1945)}]{Rice1945}%
  \BibitemOpen
  \bibfield  {author} {\bibinfo {author} {\bibfnamefont {S.}~\bibnamefont
  {Rice}},\ }\bibfield  {title} {\bibinfo {title} {Mathematical analysis of
  random noise},\ }\href@noop {} {\bibfield  {journal} {\bibinfo  {journal}
  {Bell Syst. Tech. J.}\ }\textbf {\bibinfo {volume} {24}},\ \bibinfo {pages}
  {46} (\bibinfo {year} {1945})}\BibitemShut {NoStop}%
\bibitem [{\citenamefont {Longuet-Higgins}(1952)}]{Higgins1952}%
  \BibitemOpen
  \bibfield  {author} {\bibinfo {author} {\bibfnamefont {M.}~\bibnamefont
  {Longuet-Higgins}},\ }\bibfield  {title} {\bibinfo {title} {On the
  statistical distribution of the heights of sea waves},\ }\href@noop {}
  {\bibfield  {journal} {\bibinfo  {journal} {Journal of Marine Research}\
  }\textbf {\bibinfo {volume} {11}},\ \bibinfo {pages} {245} (\bibinfo {year}
  {1952})}\BibitemShut {NoStop}%
\bibitem [{\citenamefont {Forristall}(1978)}]{Forristall1978}%
  \BibitemOpen
  \bibfield  {author} {\bibinfo {author} {\bibfnamefont {G.}~\bibnamefont
  {Forristall}},\ }\bibfield  {title} {\bibinfo {title} {On the distributions
  of wave heights in a storm},\ }\href@noop {} {\bibfield  {journal} {\bibinfo
  {journal} {J. Geophys. Res.}\ }\textbf {\bibinfo {volume} {83}},\ \bibinfo
  {pages} {2353} (\bibinfo {year} {1978})}\BibitemShut {NoStop}%
\bibitem [{\citenamefont {Tayfun}(1980)}]{Tayfun1980}%
  \BibitemOpen
  \bibfield  {author} {\bibinfo {author} {\bibfnamefont {M.~A.}\ \bibnamefont
  {Tayfun}},\ }\bibfield  {title} {\bibinfo {title} {Narrow-band nonlinear sea
  waves},\ }\href@noop {} {\bibfield  {journal} {\bibinfo  {journal} {J.
  Geophys. Res.}\ }\textbf {\bibinfo {volume} {85}},\ \bibinfo {pages} {1548}
  (\bibinfo {year} {1980})}\BibitemShut {NoStop}%
\bibitem [{\citenamefont {Karmpadakis}\ \emph {et~al.}(2020)\citenamefont
  {Karmpadakis}, \citenamefont {Swan},\ and\ \citenamefont
  {Christou}}]{Ewans2020}%
  \BibitemOpen
  \bibfield  {author} {\bibinfo {author} {\bibfnamefont {I.}~\bibnamefont
  {Karmpadakis}}, \bibinfo {author} {\bibfnamefont {C.}~\bibnamefont {Swan}},\
  and\ \bibinfo {author} {\bibfnamefont {M.}~\bibnamefont {Christou}},\
  }\bibfield  {title} {\bibinfo {title} {Assessment of wave height
  distributions using an extensive field database},\ }\href@noop {} {\bibfield
  {journal} {\bibinfo  {journal} {Coastal Eng.}\ }\textbf {\bibinfo {volume}
  {157}} (\bibinfo {year} {2020})}\BibitemShut {NoStop}%
\bibitem [{\citenamefont {Teutsch}\ \emph {et~al.}(2020)\citenamefont
  {Teutsch}, \citenamefont {Weisse}, \citenamefont {Moeller},\ and\
  \citenamefont {Krueger}}]{Teutsch2020}%
  \BibitemOpen
  \bibfield  {author} {\bibinfo {author} {\bibfnamefont {I.}~\bibnamefont
  {Teutsch}}, \bibinfo {author} {\bibfnamefont {R.}~\bibnamefont {Weisse}},
  \bibinfo {author} {\bibfnamefont {J.}~\bibnamefont {Moeller}},\ and\ \bibinfo
  {author} {\bibfnamefont {O.}~\bibnamefont {Krueger}},\ }\bibfield  {title}
  {\bibinfo {title} {A statistical analysis of rogue waves in the southern
  north sea},\ }\href@noop {} {\bibfield  {journal} {\bibinfo  {journal}
  {Natural Hazards and Earth System Sciences}\ }\textbf {\bibinfo {volume}
  {20}},\ \bibinfo {pages} {2665} (\bibinfo {year} {2020})}\BibitemShut
  {NoStop}%
\bibitem [{\citenamefont {Longuet-Higgins}(1963)}]{Higgins1963}%
  \BibitemOpen
  \bibfield  {author} {\bibinfo {author} {\bibfnamefont {M.}~\bibnamefont
  {Longuet-Higgins}},\ }\bibfield  {title} {\bibinfo {title} {The effect of
  non-linearities on statistical distributions in the theory of sea waves},\
  }\href@noop {} {\bibfield  {journal} {\bibinfo  {journal} {J. Fluid Mech.}\
  }\textbf {\bibinfo {volume} {17}},\ \bibinfo {pages} {459} (\bibinfo {year}
  {1963})}\BibitemShut {NoStop}%
\bibitem [{\citenamefont {Tayfun}\ and\ \citenamefont
  {Alkhalidi}(2020)}]{Tayfun2020}%
  \BibitemOpen
  \bibfield  {author} {\bibinfo {author} {\bibfnamefont {M.~A.}\ \bibnamefont
  {Tayfun}}\ and\ \bibinfo {author} {\bibfnamefont {M.~A.}\ \bibnamefont
  {Alkhalidi}},\ }\bibfield  {title} {\bibinfo {title} {Distribution of
  sea-surface elevations in intermediate and shallow water depths},\
  }\href@noop {} {\bibfield  {journal} {\bibinfo  {journal} {Coastal Eng.}\
  }\textbf {\bibinfo {volume} {157}} (\bibinfo {year} {2020})}\BibitemShut
  {NoStop}%
\bibitem [{\citenamefont {Bitner}(1980)}]{Bitner1980}%
  \BibitemOpen
  \bibfield  {author} {\bibinfo {author} {\bibfnamefont {E.~M.}\ \bibnamefont
  {Bitner}},\ }\bibfield  {title} {\bibinfo {title} {Non-linear effects of the
  statistical model of shallow-water wind waves},\ }\href@noop {} {\bibfield
  {journal} {\bibinfo  {journal} {Applied Ocean Research}\ }\textbf {\bibinfo
  {volume} {2}},\ \bibinfo {pages} {63} (\bibinfo {year} {1980})}\BibitemShut
  {NoStop}%
\bibitem [{\citenamefont {Tayfun}(1990)}]{Tayfun1990}%
  \BibitemOpen
  \bibfield  {author} {\bibinfo {author} {\bibfnamefont {M.}~\bibnamefont
  {Tayfun}},\ }\bibfield  {title} {\bibinfo {title} {Distribution of large wave
  heights},\ }\href@noop {} {\bibfield  {journal} {\bibinfo  {journal} {J.
  Waterway, Port, Coastal Ocean Eng.}\ }\textbf {\bibinfo {volume} {116}},\
  \bibinfo {pages} {686} (\bibinfo {year} {1990})}\BibitemShut {NoStop}%
\bibitem [{\citenamefont {Marthinsen}(1992)}]{Marthinsen1992}%
  \BibitemOpen
  \bibfield  {author} {\bibinfo {author} {\bibfnamefont {T.}~\bibnamefont
  {Marthinsen}},\ }\bibfield  {title} {\bibinfo {title} {On the statistics of
  irregular second-order waves},\ }\href@noop {} {\bibfield  {journal}
  {\bibinfo  {journal} {Report No. RMS-11}\ } (\bibinfo {year}
  {1992})}\BibitemShut {NoStop}%
\bibitem [{\citenamefont {Mori}\ and\ \citenamefont
  {Janssen}(2006)}]{Janssen2006a}%
  \BibitemOpen
  \bibfield  {author} {\bibinfo {author} {\bibfnamefont {N.}~\bibnamefont
  {Mori}}\ and\ \bibinfo {author} {\bibfnamefont {P.}~\bibnamefont {Janssen}},\
  }\bibfield  {title} {\bibinfo {title} {On kurtosis and occurrence probability
  of freak waves},\ }\href@noop {} {\bibfield  {journal} {\bibinfo  {journal}
  {J. Phys. Oceanogr.}\ }\textbf {\bibinfo {volume} {36}},\ \bibinfo {pages}
  {1471} (\bibinfo {year} {2006})}\BibitemShut {NoStop}%
\bibitem [{\citenamefont {Trulsen}\ \emph {et~al.}(2012)\citenamefont
  {Trulsen}, \citenamefont {Zeng},\ and\ \citenamefont
  {Gramstad}}]{Trulsen2012}%
  \BibitemOpen
  \bibfield  {author} {\bibinfo {author} {\bibfnamefont {K.}~\bibnamefont
  {Trulsen}}, \bibinfo {author} {\bibfnamefont {H.}~\bibnamefont {Zeng}},\ and\
  \bibinfo {author} {\bibfnamefont {O.}~\bibnamefont {Gramstad}},\ }\bibfield
  {title} {\bibinfo {title} {Laboratory evidence of freak waves provoked by
  non-uniform bathymetry},\ }\href@noop {} {\bibfield  {journal} {\bibinfo
  {journal} {Phys. Fluids}\ }\textbf {\bibinfo {volume} {24}} (\bibinfo {year}
  {2012})}\BibitemShut {NoStop}%
\bibitem [{\citenamefont {Raust\o{}l}(2014)}]{Raustol2014}%
  \BibitemOpen
  \bibfield  {author} {\bibinfo {author} {\bibfnamefont {A.}~\bibnamefont
  {Raust\o{}l}},\ }\bibfield  {title} {\bibinfo {title} {Freake b\o{}lger over
  variabelt dyp},\ }\href@noop {} {\bibfield  {journal} {\bibinfo  {journal}
  {Master's thesis, University of Oslo}\ } (\bibinfo {year}
  {2014})}\BibitemShut {NoStop}%
\bibitem [{\citenamefont {Ma}\ \emph {et~al.}(2015)\citenamefont {Ma},
  \citenamefont {Ma},\ and\ \citenamefont {Dong}}]{Ma2015}%
  \BibitemOpen
  \bibfield  {author} {\bibinfo {author} {\bibfnamefont {Y.-X.}\ \bibnamefont
  {Ma}}, \bibinfo {author} {\bibfnamefont {X.-Z.}\ \bibnamefont {Ma}},\ and\
  \bibinfo {author} {\bibfnamefont {G.-H.}\ \bibnamefont {Dong}},\ }\bibfield
  {title} {\bibinfo {title} {Variations of statistics for random waves
  propagating over a bar},\ }\href@noop {} {\bibfield  {journal} {\bibinfo
  {journal} {Journal of Marine Science and Technology (Taiwan)}\ }\textbf
  {\bibinfo {volume} {23}},\ \bibinfo {pages} {864} (\bibinfo {year}
  {2015})}\BibitemShut {NoStop}%
\bibitem [{\citenamefont {Ducrozet}\ and\ \citenamefont
  {Gouin}(2017)}]{Ducrozet2017}%
  \BibitemOpen
  \bibfield  {author} {\bibinfo {author} {\bibfnamefont {G.}~\bibnamefont
  {Ducrozet}}\ and\ \bibinfo {author} {\bibfnamefont {M.}~\bibnamefont
  {Gouin}},\ }\bibfield  {title} {\bibinfo {title} {Influence of varying
  bathymetry in rogue wave occurrence within unidirectional and directional
  sea-states},\ }\href@noop {} {\bibfield  {journal} {\bibinfo  {journal}
  {Journal of Ocean Engineering and Marine Energy}\ }\textbf {\bibinfo {volume}
  {3}} (\bibinfo {year} {2017})}\BibitemShut {NoStop}%
\bibitem [{\citenamefont {Bolles}\ \emph {et~al.}(2019)\citenamefont {Bolles},
  \citenamefont {Speer},\ and\ \citenamefont {Moore}}]{Bolles2019}%
  \BibitemOpen
  \bibfield  {author} {\bibinfo {author} {\bibfnamefont {C.}~\bibnamefont
  {Bolles}}, \bibinfo {author} {\bibfnamefont {K.}~\bibnamefont {Speer}},\ and\
  \bibinfo {author} {\bibfnamefont {M.}~\bibnamefont {Moore}},\ }\bibfield
  {title} {\bibinfo {title} {Anomalous wave statistics induced by abrupt depth
  change},\ }\href@noop {} {\bibfield  {journal} {\bibinfo  {journal} {Physical
  Review Fluids}\ }\textbf {\bibinfo {volume} {4}} (\bibinfo {year}
  {2019})}\BibitemShut {NoStop}%
\bibitem [{\citenamefont {Zhang}\ \emph {et~al.}(2019)\citenamefont {Zhang},
  \citenamefont {Benoit}, \citenamefont {Kimmoun}, \citenamefont {Chabchoub},\
  and\ \citenamefont {Hsu}}]{Chabchoub2019}%
  \BibitemOpen
  \bibfield  {author} {\bibinfo {author} {\bibfnamefont {J.}~\bibnamefont
  {Zhang}}, \bibinfo {author} {\bibfnamefont {M.}~\bibnamefont {Benoit}},
  \bibinfo {author} {\bibfnamefont {O.}~\bibnamefont {Kimmoun}}, \bibinfo
  {author} {\bibfnamefont {A.}~\bibnamefont {Chabchoub}},\ and\ \bibinfo
  {author} {\bibfnamefont {H.-C.}\ \bibnamefont {Hsu}},\ }\bibfield  {title}
  {\bibinfo {title} {Statistics of extreme waves in coastal waters: Large scale
  experiments and advanced numerical simulations},\ }\href@noop {} {\bibfield
  {journal} {\bibinfo  {journal} {Fluids}\ }\textbf {\bibinfo {volume} {4}}
  (\bibinfo {year} {2019})}\BibitemShut {NoStop}%
\bibitem [{\citenamefont {Li}\ \emph {et~al.}(2021{\natexlab{a}})\citenamefont
  {Li}, \citenamefont {Zheng}, \citenamefont {Lin}, \citenamefont {Adcock},\
  and\ \citenamefont {Van Den~Bremer}}]{Adcock2021a}%
  \BibitemOpen
  \bibfield  {author} {\bibinfo {author} {\bibfnamefont {Y.}~\bibnamefont
  {Li}}, \bibinfo {author} {\bibfnamefont {Y.}~\bibnamefont {Zheng}}, \bibinfo
  {author} {\bibfnamefont {Z.}~\bibnamefont {Lin}}, \bibinfo {author}
  {\bibfnamefont {T.~A.}\ \bibnamefont {Adcock}},\ and\ \bibinfo {author}
  {\bibfnamefont {T.}~\bibnamefont {Van Den~Bremer}},\ }\bibfield  {title}
  {\bibinfo {title} {Surface wavepackets subject to an abrupt depth change.
  part 1: Second-order theory},\ }\href@noop {} {\bibfield  {journal} {\bibinfo
   {journal} {J. Fluid Mech.}\ }\textbf {\bibinfo {volume} {915}},\ \bibinfo
  {pages} {A71} (\bibinfo {year} {2021}{\natexlab{a}})}\BibitemShut {NoStop}%
\bibitem [{\citenamefont {Trulsen}\ \emph {et~al.}(2020)\citenamefont
  {Trulsen}, \citenamefont {Raust\o{}l}, \citenamefont {Jorde},\ and\
  \citenamefont {Rye}}]{Trulsen2020}%
  \BibitemOpen
  \bibfield  {author} {\bibinfo {author} {\bibfnamefont {K.}~\bibnamefont
  {Trulsen}}, \bibinfo {author} {\bibfnamefont {A.}~\bibnamefont {Raust\o{}l}},
  \bibinfo {author} {\bibfnamefont {S.}~\bibnamefont {Jorde}},\ and\ \bibinfo
  {author} {\bibfnamefont {L.}~\bibnamefont {Rye}},\ }\bibfield  {title}
  {\bibinfo {title} {Extreme wave statistics of long-crested irregular waves
  over a shoal},\ }\href@noop {} {\bibfield  {journal} {\bibinfo  {journal} {J.
  Fluid Mech.}\ }\textbf {\bibinfo {volume} {882}} (\bibinfo {year}
  {2020})}\BibitemShut {NoStop}%
\bibitem [{\citenamefont {Mendes}\ and\ \citenamefont
  {Kasparian}(2022)}]{Mendes2022b}%
  \BibitemOpen
  \bibfield  {author} {\bibinfo {author} {\bibfnamefont {S.}~\bibnamefont
  {Mendes}}\ and\ \bibinfo {author} {\bibfnamefont {J.}~\bibnamefont
  {Kasparian}},\ }\bibfield  {title} {\bibinfo {title} {Saturation of rogue
  wave amplification over steep shoals},\ }\href@noop {} {\bibfield  {journal}
  {\bibinfo  {journal} {Phys. Rev. E}\ }\textbf {\bibinfo {volume} {106}},\
  \bibinfo {pages} {065101} (\bibinfo {year} {2022})}\BibitemShut {NoStop}%
\bibitem [{\citenamefont {Moore}\ \emph {et~al.}(2020)\citenamefont {Moore},
  \citenamefont {Bolles}, \citenamefont {Majda},\ and\ \citenamefont
  {Qi}}]{Moore2020}%
  \BibitemOpen
  \bibfield  {author} {\bibinfo {author} {\bibfnamefont {N.}~\bibnamefont
  {Moore}}, \bibinfo {author} {\bibfnamefont {C.}~\bibnamefont {Bolles}},
  \bibinfo {author} {\bibfnamefont {A.}~\bibnamefont {Majda}},\ and\ \bibinfo
  {author} {\bibfnamefont {D.}~\bibnamefont {Qi}},\ }\bibfield  {title}
  {\bibinfo {title} {Anomalous waves triggered by abrupt depth changes:
  Laboratory experiments and truncated kdv statistical mechanics},\ }\href@noop
  {} {\bibfield  {journal} {\bibinfo  {journal} {Journal of Nonlinear Science}\
  }\textbf {\bibinfo {volume} {30}},\ \bibinfo {pages} {3235} (\bibinfo {year}
  {2020})}\BibitemShut {NoStop}%
\bibitem [{\citenamefont {Li}\ \emph {et~al.}(2021{\natexlab{b}})\citenamefont
  {Li}, \citenamefont {Draycott}, \citenamefont {Zheng}, \citenamefont {Lin},
  \citenamefont {Adcock},\ and\ \citenamefont {Van Den~Bremer}}]{Adcock2021c}%
  \BibitemOpen
  \bibfield  {author} {\bibinfo {author} {\bibfnamefont {Y.}~\bibnamefont
  {Li}}, \bibinfo {author} {\bibfnamefont {S.}~\bibnamefont {Draycott}},
  \bibinfo {author} {\bibfnamefont {Y.}~\bibnamefont {Zheng}}, \bibinfo
  {author} {\bibfnamefont {Z.}~\bibnamefont {Lin}}, \bibinfo {author}
  {\bibfnamefont {T.}~\bibnamefont {Adcock}},\ and\ \bibinfo {author}
  {\bibfnamefont {T.}~\bibnamefont {Van Den~Bremer}},\ }\bibfield  {title}
  {\bibinfo {title} {Why rogue waves occur atop abrupt depth transitions},\
  }\href@noop {} {\bibfield  {journal} {\bibinfo  {journal} {Journal of Fluid
  Mechanics}\ }\textbf {\bibinfo {volume} {919}},\ \bibinfo {pages} {R5}
  (\bibinfo {year} {2021}{\natexlab{b}})}\BibitemShut {NoStop}%
\bibitem [{\citenamefont {Mendes}\ \emph {et~al.}(2022)\citenamefont {Mendes},
  \citenamefont {Scotti}, \citenamefont {Brunetti},\ and\ \citenamefont
  {Kasparian}}]{Mendes2021b}%
  \BibitemOpen
  \bibfield  {author} {\bibinfo {author} {\bibfnamefont {S.}~\bibnamefont
  {Mendes}}, \bibinfo {author} {\bibfnamefont {A.}~\bibnamefont {Scotti}},
  \bibinfo {author} {\bibfnamefont {M.}~\bibnamefont {Brunetti}},\ and\
  \bibinfo {author} {\bibfnamefont {J.}~\bibnamefont {Kasparian}},\ }\bibfield
  {title} {\bibinfo {title} {Non-homogeneous model of rogue wave probability
  evolution over a shoal},\ }\href@noop {} {\bibfield  {journal} {\bibinfo
  {journal} {J. Fluid Mech.}\ }\textbf {\bibinfo {volume} {939}},\ \bibinfo
  {pages} {A25} (\bibinfo {year} {2022})}\BibitemShut {NoStop}%
\bibitem [{\citenamefont {Janssen}\ and\ \citenamefont
  {Bidlot}(2009)}]{Janssen2009}%
  \BibitemOpen
  \bibfield  {author} {\bibinfo {author} {\bibfnamefont {P.~E.~M.}\
  \bibnamefont {Janssen}}\ and\ \bibinfo {author} {\bibfnamefont {J.-R.}\
  \bibnamefont {Bidlot}},\ }\href@noop {} {\emph {\bibinfo {title} {On the
  extension of the freak wave warning system and its verification}}}\ (\bibinfo
   {publisher} {European Centre for Medium-Range Weather Forecasts Reading,
  UK},\ \bibinfo {year} {2009})\BibitemShut {NoStop}%
\bibitem [{\citenamefont {Casas-Prat}\ and\ \citenamefont
  {Holthuijsen}(2010)}]{Holthuijsen2010}%
  \BibitemOpen
  \bibfield  {author} {\bibinfo {author} {\bibfnamefont {M.}~\bibnamefont
  {Casas-Prat}}\ and\ \bibinfo {author} {\bibfnamefont {L.}~\bibnamefont
  {Holthuijsen}},\ }\bibfield  {title} {\bibinfo {title} {Short-term statistics
  of waves observed in deep water},\ }\href@noop {} {\bibfield  {journal}
  {\bibinfo  {journal} {J. Geophys. Res. Oceans}\ }\textbf {\bibinfo {volume}
  {115}} (\bibinfo {year} {2010})}\BibitemShut {NoStop}%
\bibitem [{\citenamefont {Mendes}\ \emph {et~al.}(2021)\citenamefont {Mendes},
  \citenamefont {Scotti},\ and\ \citenamefont {Stansell}}]{Mendes2021a}%
  \BibitemOpen
  \bibfield  {author} {\bibinfo {author} {\bibfnamefont {S.}~\bibnamefont
  {Mendes}}, \bibinfo {author} {\bibfnamefont {A.}~\bibnamefont {Scotti}},\
  and\ \bibinfo {author} {\bibfnamefont {P.}~\bibnamefont {Stansell}},\
  }\bibfield  {title} {\bibinfo {title} {On the physical constraints for the
  exceeding probability of deep water rogue waves},\ }\href@noop {} {\bibfield
  {journal} {\bibinfo  {journal} {Appl. Ocean Res.}\ }\textbf {\bibinfo
  {volume} {108}},\ \bibinfo {pages} {102402} (\bibinfo {year}
  {2021})}\BibitemShut {NoStop}%
\bibitem [{\citenamefont {Goda}(1983)}]{Goda1983}%
  \BibitemOpen
  \bibfield  {author} {\bibinfo {author} {\bibfnamefont {Y.}~\bibnamefont
  {Goda}},\ }\bibfield  {title} {\bibinfo {title} {A unified nonlinearity
  parameter of water waves},\ }\href@noop {} {\bibfield  {journal} {\bibinfo
  {journal} {Rept. Port and Harbour Res. Inst.}\ }\textbf {\bibinfo {volume}
  {22 (3)}},\ \bibinfo {pages} {3} (\bibinfo {year} {1983})}\BibitemShut
  {NoStop}%
\bibitem [{\citenamefont {Battjes}(1974)}]{Battjes1974}%
  \BibitemOpen
  \bibfield  {author} {\bibinfo {author} {\bibfnamefont {J.}~\bibnamefont
  {Battjes}},\ }\bibfield  {title} {\bibinfo {title} {Surf similarity},\
  }\href@noop {} {\bibfield  {journal} {\bibinfo  {journal} {Coastal
  Engineering Proceedings}\ }\textbf {\bibinfo {volume} {1}},\ \bibinfo {pages}
  {26} (\bibinfo {year} {1974})}\BibitemShut {NoStop}%
\bibitem [{\citenamefont {Kimmoun}\ \emph {et~al.}(2021)\citenamefont
  {Kimmoun}, \citenamefont {Hsu}, \citenamefont {Hoffmann},\ and\ \citenamefont
  {Chabchoub}}]{Chabchoub2021}%
  \BibitemOpen
  \bibfield  {author} {\bibinfo {author} {\bibfnamefont {O.}~\bibnamefont
  {Kimmoun}}, \bibinfo {author} {\bibfnamefont {H.-C.}\ \bibnamefont {Hsu}},
  \bibinfo {author} {\bibfnamefont {N.}~\bibnamefont {Hoffmann}},\ and\
  \bibinfo {author} {\bibfnamefont {A.}~\bibnamefont {Chabchoub}},\ }\bibfield
  {title} {\bibinfo {title} {Experiments on uni-directional and nonlinear wave
  group shoaling},\ }\href@noop {} {\bibfield  {journal} {\bibinfo  {journal}
  {Ocean Dynamics}\ } (\bibinfo {year} {2021})}\BibitemShut {NoStop}%
\bibitem [{\citenamefont {Glukhovskiy}(1966)}]{Glukhovskii1966}%
  \BibitemOpen
  \bibfield  {author} {\bibinfo {author} {\bibfnamefont {B.~K.}\ \bibnamefont
  {Glukhovskiy}},\ }\bibfield  {title} {\bibinfo {title} {Investigation of sea
  wind waves (in russian)},\ }\href@noop {} {\bibfield  {journal} {\bibinfo
  {journal} {Gidrometeoizdat, Leningrad,}\ ,\ \bibinfo {pages} {283}} (\bibinfo
  {year} {1966})}\BibitemShut {NoStop}%
\bibitem [{\citenamefont {Karmpadakis}\ \emph {et~al.}(2022)\citenamefont
  {Karmpadakis}, \citenamefont {Swan},\ and\ \citenamefont
  {Christou}}]{Karmpadakis2022}%
  \BibitemOpen
  \bibfield  {author} {\bibinfo {author} {\bibfnamefont {I.}~\bibnamefont
  {Karmpadakis}}, \bibinfo {author} {\bibfnamefont {C.}~\bibnamefont {Swan}},\
  and\ \bibinfo {author} {\bibfnamefont {M.}~\bibnamefont {Christou}},\
  }\bibfield  {title} {\bibinfo {title} {A new wave height distribution for
  intermediate and shallow water depths},\ }\href@noop {} {\bibfield  {journal}
  {\bibinfo  {journal} {Coastal Engineering}\ }\textbf {\bibinfo {volume}
  {175}},\ \bibinfo {pages} {104130} (\bibinfo {year} {2022})}\BibitemShut
  {NoStop}%
\bibitem [{\citenamefont {Mori}\ and\ \citenamefont
  {Kobayashi}(1998)}]{Mori1998}%
  \BibitemOpen
  \bibfield  {author} {\bibinfo {author} {\bibfnamefont {N.}~\bibnamefont
  {Mori}}\ and\ \bibinfo {author} {\bibfnamefont {N.}~\bibnamefont
  {Kobayashi}},\ }\bibfield  {title} {\bibinfo {title} {Nonlinear distribution
  of neashore free surface and velocity},\ }in\ \href@noop {} {\emph {\bibinfo
  {booktitle} {Coastal Engineering 1998}}}\ (\bibinfo {year} {1998})\ pp.\
  \bibinfo {pages} {189--202}\BibitemShut {NoStop}%
\bibitem [{\citenamefont {Mori}\ and\ \citenamefont
  {Yasuda}(2002)}]{Mori2002b}%
  \BibitemOpen
  \bibfield  {author} {\bibinfo {author} {\bibfnamefont {N.}~\bibnamefont
  {Mori}}\ and\ \bibinfo {author} {\bibfnamefont {T.}~\bibnamefont {Yasuda}},\
  }\bibfield  {title} {\bibinfo {title} {A weakly non-gaussian model of wave
  height distribution random wave train},\ }\href@noop {} {\bibfield  {journal}
  {\bibinfo  {journal} {Ocean Eng.}\ }\textbf {\bibinfo {volume} {29}},\
  \bibinfo {pages} {1219–1231} (\bibinfo {year} {2002})}\BibitemShut
  {NoStop}%
\bibitem [{\citenamefont {Mendes}(2020)}]{BMendes2020}%
  \BibitemOpen
  \bibfield  {author} {\bibinfo {author} {\bibfnamefont {S.}~\bibnamefont
  {Mendes}},\ }\bibfield  {title} {\bibinfo {title} {On the statistics of
  oceanic rogue waves in finite depth: Exceeding probabilities, physical
  constraints and extreme value theory},\ }\href@noop {} {\bibfield  {journal}
  {\bibinfo  {journal} {UNC Chapel Hill PhD Thesis}\ } (\bibinfo {year}
  {2020})}\BibitemShut {NoStop}%
\bibitem [{\citenamefont {Longuet-Higgins}(1975)}]{Higgins1975}%
  \BibitemOpen
  \bibfield  {author} {\bibinfo {author} {\bibfnamefont {M.~S.}\ \bibnamefont
  {Longuet-Higgins}},\ }\bibfield  {title} {\bibinfo {title} {On the joint
  distribution of the periods and amplitudes of sea waves},\ }\href@noop {}
  {\bibfield  {journal} {\bibinfo  {journal} {J. Geophys. Res.}\ }\textbf
  {\bibinfo {volume} {80}},\ \bibinfo {pages} {2688} (\bibinfo {year}
  {1975})}\BibitemShut {NoStop}%
\bibitem [{\citenamefont {Tayfun}(2006)}]{Tayfun2006}%
  \BibitemOpen
  \bibfield  {author} {\bibinfo {author} {\bibfnamefont {M.}~\bibnamefont
  {Tayfun}},\ }\bibfield  {title} {\bibinfo {title} {Statistics of nonlinear
  wave crests and groups},\ }\href@noop {} {\bibfield  {journal} {\bibinfo
  {journal} {Ocean Eng.}\ }\textbf {\bibinfo {volume} {33}},\ \bibinfo {pages}
  {1589} (\bibinfo {year} {2006})}\BibitemShut {NoStop}%
\bibitem [{\citenamefont {Stansell}(2004)}]{Stansell2004}%
  \BibitemOpen
  \bibfield  {author} {\bibinfo {author} {\bibfnamefont {P.}~\bibnamefont
  {Stansell}},\ }\bibfield  {title} {\bibinfo {title} {Distribution of freak
  wave heights measured in the north sea},\ }\href@noop {} {\bibfield
  {journal} {\bibinfo  {journal} {Appl. Ocean Res.}\ }\textbf {\bibinfo
  {volume} {26}},\ \bibinfo {pages} {35} (\bibinfo {year} {2004})}\BibitemShut
  {NoStop}%
\bibitem [{\citenamefont {Stansell}(2005)}]{Stansell2005}%
  \BibitemOpen
  \bibfield  {author} {\bibinfo {author} {\bibfnamefont {P.}~\bibnamefont
  {Stansell}},\ }\bibfield  {title} {\bibinfo {title} {Distributions of extreme
  wave, crest and trough heights measured in the north sea},\ }\href@noop {}
  {\bibfield  {journal} {\bibinfo  {journal} {Ocean Eng.}\ }\textbf {\bibinfo
  {volume} {32}},\ \bibinfo {pages} {1015} (\bibinfo {year}
  {2005})}\BibitemShut {NoStop}%
\bibitem [{\citenamefont {Linfoot}\ \emph {et~al.}(2000)\citenamefont
  {Linfoot}, \citenamefont {Stansell},\ and\ \citenamefont
  {Wolfram}}]{Stansell2000}%
  \BibitemOpen
  \bibfield  {author} {\bibinfo {author} {\bibfnamefont {B.}~\bibnamefont
  {Linfoot}}, \bibinfo {author} {\bibfnamefont {P.}~\bibnamefont {Stansell}},\
  and\ \bibinfo {author} {\bibfnamefont {J.}~\bibnamefont {Wolfram}},\
  }\bibfield  {title} {\bibinfo {title} {On the characteristics of storm
  waves},\ }\href@noop {} {\bibfield  {journal} {\bibinfo  {journal}
  {Proceedings of the International Offshore and Polar Engineering Conference}\
  }\textbf {\bibinfo {volume} {3}},\ \bibinfo {pages} {74} (\bibinfo {year}
  {2000})}\BibitemShut {NoStop}%
\bibitem [{\citenamefont {Doeleman}(2021)}]{Doeleman2021}%
  \BibitemOpen
  \bibfield  {author} {\bibinfo {author} {\bibfnamefont {M.~W.}\ \bibnamefont
  {Doeleman}},\ }\bibfield  {title} {\bibinfo {title} {Rogue waves in the dutch
  north sea},\ }\href@noop {} {\bibfield  {journal} {\bibinfo  {journal}
  {Master's thesis, TU Delft}\ } (\bibinfo {year} {2021})}\BibitemShut
  {NoStop}%
\bibitem [{\citenamefont {Janssen}(2017)}]{Janssen2017}%
  \BibitemOpen
  \bibfield  {author} {\bibinfo {author} {\bibfnamefont {P.~A. E.~M.}\
  \bibnamefont {Janssen}},\ }\href@noop {} {\emph {\bibinfo {title}
  {Shallow-water version of the freak wave warning system}}}\ (\bibinfo
  {publisher} {European Centre for Medium Range Weather Forecasts},\ \bibinfo
  {year} {2017})\BibitemShut {NoStop}%
\bibitem [{\citenamefont {Borgman}(1973)}]{Borgman1973}%
  \BibitemOpen
  \bibfield  {author} {\bibinfo {author} {\bibfnamefont {L.~E.}\ \bibnamefont
  {Borgman}},\ }\bibfield  {title} {\bibinfo {title} {Probabilities for highest
  wave in hurricane},\ }\href@noop {} {\bibfield  {journal} {\bibinfo
  {journal} {Journal of the Waterways, Harbors and Coastal Engineering
  Division}\ }\textbf {\bibinfo {volume} {99}},\ \bibinfo {pages} {185}
  (\bibinfo {year} {1973})}\BibitemShut {NoStop}%
\bibitem [{\citenamefont {Muir}\ and\ \citenamefont
  {El-Shaarawi}(1986)}]{El-Shaarawi1986}%
  \BibitemOpen
  \bibfield  {author} {\bibinfo {author} {\bibfnamefont {L.~R.}\ \bibnamefont
  {Muir}}\ and\ \bibinfo {author} {\bibfnamefont {A.}~\bibnamefont
  {El-Shaarawi}},\ }\bibfield  {title} {\bibinfo {title} {On the calculation of
  extreme wave heights: a review},\ }\href@noop {} {\bibfield  {journal}
  {\bibinfo  {journal} {Ocean Engineering}\ }\textbf {\bibinfo {volume} {13}},\
  \bibinfo {pages} {93} (\bibinfo {year} {1986})}\BibitemShut {NoStop}%
\bibitem [{\citenamefont {Zhang}\ \emph {et~al.}(2023)\citenamefont {Zhang},
  \citenamefont {Ma}, \citenamefont {Tan}, \citenamefont {Dong},\ and\
  \citenamefont {Benoit}}]{Benoit2023}%
  \BibitemOpen
  \bibfield  {author} {\bibinfo {author} {\bibfnamefont {J.}~\bibnamefont
  {Zhang}}, \bibinfo {author} {\bibfnamefont {Y.}~\bibnamefont {Ma}}, \bibinfo
  {author} {\bibfnamefont {T.}~\bibnamefont {Tan}}, \bibinfo {author}
  {\bibfnamefont {G.}~\bibnamefont {Dong}},\ and\ \bibinfo {author}
  {\bibfnamefont {M.}~\bibnamefont {Benoit}},\ }\bibfield  {title} {\bibinfo
  {title} {Enhanced extreme wave statistics of irregular waves due to
  accelerating following current over a submerged bar},\ }\href@noop {}
  {\bibfield  {journal} {\bibinfo  {journal} {J. Fluid Mech.}\ }\textbf
  {\bibinfo {volume} {954}},\ \bibinfo {pages} {A50} (\bibinfo {year}
  {2023})}\BibitemShut {NoStop}%
\bibitem [{\citenamefont {Figueras}(2010)}]{Figueras2010}%
  \BibitemOpen
  \bibfield  {author} {\bibinfo {author} {\bibfnamefont {A.}~\bibnamefont
  {Figueras}},\ }\bibfield  {title} {\bibinfo {title} {Estimation of available
  wave power in the near shore area around hanstholm harbor},\ }\href@noop {}
  {\bibfield  {journal} {\bibinfo  {journal} {Project of Special Thesis,
  Universidad Politecnica de Catalunya}\ } (\bibinfo {year}
  {2010})}\BibitemShut {NoStop}%
\bibitem [{\citenamefont {Holthuijsen}(2007)}]{Holthuijsen2007}%
  \BibitemOpen
  \bibfield  {author} {\bibinfo {author} {\bibfnamefont {L.~H.}\ \bibnamefont
  {Holthuijsen}},\ }\href@noop {} {\emph {\bibinfo {title} {Waves in Oceanic
  and Coastal Waters}}}\ (\bibinfo  {publisher} {Cambridge University Press},\
  \bibinfo {year} {2007})\BibitemShut {NoStop}%
\bibitem [{\citenamefont {Goda}(1975)}]{Goda1975}%
  \BibitemOpen
  \bibfield  {author} {\bibinfo {author} {\bibfnamefont {Y.}~\bibnamefont
  {Goda}},\ }\bibfield  {title} {\bibinfo {title} {Irregular wave deformation
  in the surf zone},\ }\href@noop {} {\bibfield  {journal} {\bibinfo  {journal}
  {Coastal Engineering in Japan}\ }\textbf {\bibinfo {volume} {18}},\ \bibinfo
  {pages} {13} (\bibinfo {year} {1975})}\BibitemShut {NoStop}%
\bibitem [{\citenamefont {Goda}(2010)}]{Goda2010}%
  \BibitemOpen
  \bibfield  {author} {\bibinfo {author} {\bibfnamefont {Y.}~\bibnamefont
  {Goda}},\ }\bibfield  {title} {\bibinfo {title} {Random seas for design of
  maritime structures},\ }\href@noop {} {\bibfield  {journal} {\bibinfo
  {journal} {World Scientific}\ } (\bibinfo {year} {2010})}\BibitemShut
  {NoStop}%
\bibitem [{\citenamefont {Eagleson}(1956)}]{Eagleson1956}%
  \BibitemOpen
  \bibfield  {author} {\bibinfo {author} {\bibfnamefont {P.~S.}\ \bibnamefont
  {Eagleson}},\ }\bibfield  {title} {\bibinfo {title} {Properties of shoaling
  waves by theory and experiment},\ }\href@noop {} {\bibfield  {journal}
  {\bibinfo  {journal} {Eos, Transactions American Geophysical Union}\ }\textbf
  {\bibinfo {volume} {37}},\ \bibinfo {pages} {565} (\bibinfo {year}
  {1956})}\BibitemShut {NoStop}%
\bibitem [{\citenamefont {Walker}\ and\ \citenamefont
  {Headlam}(1983)}]{Walker1983}%
  \BibitemOpen
  \bibfield  {author} {\bibinfo {author} {\bibfnamefont {J.}~\bibnamefont
  {Walker}}\ and\ \bibinfo {author} {\bibfnamefont {J.}~\bibnamefont
  {Headlam}},\ }\bibfield  {title} {\bibinfo {title} {Engineering approach to
  nonlinear wave shoaling},\ }\href@noop {} {\bibfield  {journal} {\bibinfo
  {journal} {Proceedings of the Coastal Engineering Conference}\ }\textbf
  {\bibinfo {volume} {1}},\ \bibinfo {pages} {523} (\bibinfo {year}
  {1983})}\BibitemShut {NoStop}%
\bibitem [{\citenamefont {Srineash}\ and\ \citenamefont
  {Murali}(2018)}]{Srineash2018}%
  \BibitemOpen
  \bibfield  {author} {\bibinfo {author} {\bibfnamefont {V.}~\bibnamefont
  {Srineash}}\ and\ \bibinfo {author} {\bibfnamefont {K.}~\bibnamefont
  {Murali}},\ }\bibfield  {title} {\bibinfo {title} {Wave shoaling over a
  submerged ramp: An experimental and numerical study},\ }\href@noop {}
  {\bibfield  {journal} {\bibinfo  {journal} {Journal of Waterway, Port,
  Coastal and Ocean Engineering}\ }\textbf {\bibinfo {volume} {144}} (\bibinfo
  {year} {2018})}\BibitemShut {NoStop}%
\end{thebibliography}%

\end{document}